\documentclass[12pt]{spieman}  
\usepackage{amsmath,amsfonts,amssymb}
\usepackage{graphicx}
\usepackage{setspace}
\usepackage{tocloft}

\usepackage{threeparttable}
\usepackage{multirow}
\usepackage{siunitx}
\DeclareSIUnit{\Molar}{M}
\DeclareSIUnit{\Hg}{Hg}
\DeclareSIUnit{\yr}{yr}
\usepackage{GB_glossaries}
\usepackage{soul}
\usepackage[table]{xcolor}
\usepackage{dirtytalk}
\usepackage{wasysym}

\usepackage{lineno}

\usepackage{qting}

\selectcolormodel{RGB}
\definecolor{blu}{RGB}{ 49, 114, 174} 
\definecolor{brn}{RGB}{ 94,  75,  60} 
\definecolor{gry}{RGB}{100, 100, 105} 
\definecolor{org}{RGB}{212,  93,   0} 
\definecolor{red}{RGB}{203,  51,  59} 
\definecolor{pur}{RGB}{ 80,   7, 120} 
\definecolor{grn}{RGB}{ 86, 108,  17} 
\definecolor{lgr}{RGB}{ 98, 166,  10} 
\AtBeginDocument{\colorlet{defaultcolor}{.}}
\newcommand{\rl}[1]{\textcolor{defaultcolor}{#1}}
\newcommand{\rlA}[1]{\textcolor{defaultcolor}{#1}}
\newcommand{\rlB}[1]{\textcolor{defaultcolor}{#1}}

\definecolor{DScol1}{rgb}{1.0,0.0,0.0} 
\definecolor{DScol2}{rgb}{1.0,0.2,0.0} 
\definecolor{DScol3}{rgb}{1.0,0.4,0.0} 
\definecolor{DScol4}{rgb}{1.0,0.6,0.0} 
\definecolor{DScol5}{rgb}{1.0,0.8,0.0} 
\definecolor{DScol6}{rgb}{1.0,1.0,0.0} 
\definecolor{DScol7}{rgb}{0.8,1.0,0.0} 
\definecolor{DScol8}{rgb}{0.6,1.0,0.0} 
\definecolor{DScol9}{rgb}{0.4,1.0,0.0} 
\definecolor{DScol10}{rgb}{0.2,1.0,0.0} 
\definecolor{DScol11}{rgb}{0.0,1.0,0.0} 
\definecolor{DScol12}{rgb}{0.0,1.0,0.2} 
\definecolor{DScol13}{rgb}{0.0,1.0,0.4} 
\definecolor{DScol14}{rgb}{0.0,1.0,0.6} 
\definecolor{DScol15}{rgb}{0.0,1.0,0.8} 
\definecolor{DScol16}{rgb}{0.0,1.0,1.0} 
\definecolor{DScol17}{rgb}{0.0,0.8,1.0} 
\definecolor{DScol18}{rgb}{0.0,0.6,1.0} 
\definecolor{DScol19}{rgb}{0.0,0.4,1.0} 
\definecolor{DScol20}{rgb}{0.0,0.2,1.0} 
\definecolor{DScol21}{rgb}{0.0,0.0,1.0} 
\definecolor{DScol22}{rgb}{0.2,0.0,1.0} 
\definecolor{DScol23}{rgb}{0.4,0.0,1.0} 
\definecolor{DScol24}{rgb}{0.6,0.0,1.0} 
\definecolor{DScol25}{rgb}{0.8,0.0,1.0} 
\definecolor{DScol26}{rgb}{1.0,0.0,1.0} 
\definecolor{DScol27}{rgb}{1.0,0.0,0.8} 
\definecolor{DScol28}{rgb}{1.0,0.0,0.6} 
\definecolor{DScol29}{rgb}{1.0,0.0,0.4} 
\definecolor{DScol30}{rgb}{1.0,0.0,0.2} 

\newcommand{\ie}[1]{(\textit{i.e.}, #1)}

\title{Dual-slope imaging of cerebral hemodynamics with frequency-domain near-infrared spectroscopy}

\author{Giles~Blaney\textsuperscript{\textdagger,\textasteriskcentered}}
\author{Cristianne~Fernandez\textsuperscript{\textdagger}}
\author{Angelo~Sassaroli}
\author{Sergio~Fantini}
\affil{Tufts University, Department of Biomedical Engineering, 4 Colby Street, Medford MA, USA, 02155}

\cftpagenumbersoff{figure}
\cftpagenumbersoff{table} 
\begin{document} 
\maketitle

\begin{abstract} 
    \\
    \noindent\textbf{Significance:} This work targets the contamination of optical signals by superficial hemodynamics, which is one of the chief hurdles in non-invasive optical measurements of the human brain.
    \vspace{1ex}
    
    \noindent\textbf{Aim:} To identify optimal source-detector distances for \acrfull{DS} measurements in \acrfull{FD} \acrfull{NIRS} and demonstrate preferential sensitivity of \acrfull{DS} imaging to deeper tissue (brain) versus superficial tissue (scalp).
    \vspace{1ex}
    
    \noindent\textbf{Approach:} Theoretical studies (\textit{in-silico}) based on diffusion theory in two-layered and in homogeneous scattering media. \textit{In-vivo} demonstrations of \acrshort{DS} imaging of the human brain during visual stimulation and during systemic blood pressure oscillations.
    \vspace{1ex}
    
    \noindent\textbf{Results:} The mean distance (between the two source-detector distances needed for \acrshort{DS}) is the key factor for depth sensitivity. \textit{In-vivo} imaging of the human \textit{occipital} lobe with \acrshort{FD} \acrshort{NIRS} and a mean distance of \SI{31}{\milli\meter} indicated: \rl{(1) greater hemodynamic response to visual stimulation from \acrshort{FD} phase versus intensity, and from \acrshort{DS} versus \acrfull{SD}; (2) hemodynamics from \acrshort{FD} phase and \acrshort{DS} mainly driven by blood flow, and hemodynamics from \acrshort{SD} intensity mainly driven by blood volume.}
    \vspace{1ex}
    
    \noindent\textbf{Conclusions:} \Acrshort{DS} imaging with \acrshort{FD} \acrshort{NIRS} may suppress confounding contributions from superficial hemodynamics without relying on data at short source-detector distances. This capability can have significant implications for non-invasive optical measurements of the human brain.
\end{abstract}


{\noindent \footnotesize\textbf{\textasteriskcentered}Giles~Blaney,  \linkable{Giles.Blaney@tufts.edu}\\
\noindent \footnotesize\textbf{\textdagger}Giles~Blaney \& Cristianne~Fernandez contributed equally}

\begin{spacing}{1} 

\section{Introduction}\label{sec:intro}
Functional brain \gls{DOI} using \gls{NIRS} has seen an increase in its popularity and applications over the past \SI{30}{\yr}\cite{boasTwentyYearsFunctional2014, quaresimaFunctionalNearInfraredSpectroscopy2019, rahmanNarrativeReviewClinical2020}. During that time, \gls{fNIRS} has been demonstrated in both behavioural and social studies\cite{quaresimaFunctionalNearInfraredSpectroscopy2019} and in clinical applications\cite{rahmanNarrativeReviewClinical2020}. A large reason for the success of \gls{fNIRS} is due to its ability to spatially map brain hemodynamics and activation in specific cortical regions while being non-invasive, portable, and low-cost especially when comparing the latter two advantages to \gls{fMRI}\cite{eggebrechtQuantitativeSpatialComparison2012}. Moreover, \gls{NIRS} offers continuous monitoring of key target organs, not only at the bedside but in real-life settings\cite{rahmanNarrativeReviewClinical2020}. However, \gls{fNIRS} and \gls{DOI} still struggle with one of their largest weaknesses, a significant sensitivity to superficial, extracerebral tissue\cite{franceschiniInfluenceSuperficialLayer1998,gagnonQuantificationCorticalContribution2012,saagerMeasurementLayerlikeHemodynamic2008,funaneQuantitativeEvaluationDeep2014}. Despite the aforementioned advantages of \gls{DOI}, most techniques still preferentially measure scalp and skull hemodynamics; with only a weak contribution from the brain itself. 
\rl{Therefore, the field has continually investigated methods that seek to identify or suppress this superficial signal, and allow for more specific brain measurements\cite{zhangAdaptiveFilteringGlobal2007,saagerMeasurementLayerlikeHemodynamic2008,funaneQuantitativeEvaluationDeep2014,gagnonFurtherImprovementReducing2014,blaneyPhaseDualslopesFrequencydomain2020,veesaSignalRegressionFrequencydomain2021}.} \par

The cheapest and most common implementations of \gls{fNIRS} and \gls{DOI} have utilized \gls{CW}, methods that are most strongly affected by superficial hemodynamics. 
\rl{In \gls{FD}\cite{fantiniFrequencyDomainTechniquesCerebral2020} or \gls{TD}\cite{torricelliTimeDomainFunctional2014} techniques, the \gls{phi} or higher moments of the photon time-of-flight distribution, respectively, intrinsically provide measurements that are more specific to deep regions.}
\rl{Despite this, due to the widespread use of \gls{fNIRS}, a majority of the aforementioned techniques to determine brain's contribution to the signal are targeted toward \gls{CW} data and include measurements that are specifically sensitive to superficial hemodynamics\cite{zhangAdaptiveFilteringGlobal2007,saagerMeasurementLayerlikeHemodynamic2008,funaneQuantitativeEvaluationDeep2014,gagnonFurtherImprovementReducing2014}.}
Recently, there has been a push to implement imaging arrays using \gls{FD} or \gls{TD} \gls{NIRS} to gain \gls{phi} or higher moments information in an attempt to retrieve optical data that are intrinsically sensitive to deeper tissue\cite{sawoszMethodImproveDepth2019,doulgerakisHighdensityFunctionalDiffuse2019,blaneyPhaseDualslopesFrequencydomain2020,perkinsQuantitativeEvaluationFrequency2021,veesaSignalRegressionFrequencydomain2021}. 
\rl{Furthermore, typical implementations of \gls{DOI} utilize \gls{SD} based source-detector arrangements that consist of source-detector pairs spaced at various \glspl{rho} across the \gls{SD} sets\cite{vidal-rosasEvaluatinganewgeneration2021,eggebrechtQuantitativeSpatialComparison2012}.}
\Gls{SD} measurements are known to be largely sensitive to superficial tissue. 
\rl{To combat this problem, the combination of data collected at different \glspl{rho} or many \glspl{SD} has been used to minimize signal contributions associated with superficial tissue in some way\cite{zhangAdaptiveFilteringGlobal2007,saagerMeasurementLayerlikeHemodynamic2008,eggebrechtQuantitativeSpatialComparison2012,gagnonFurtherImprovementReducing2014,funaneQuantitativeEvaluationDeep2014,doulgerakisHighdensityFunctionalDiffuse2019,veesaSignalRegressionFrequencydomain2021,perkinsQuantitativeEvaluationFrequency2021}.}
\rl{However, it is still unclear which set of \glspl{rho} will optimally reconstruct deep tissue dynamics.}
A method that has been introduced to achieve this subtraction intrinsically is the \gls{DS}\cite{sassaroliDualslopeMethodEnhanced2019,blaneyPhaseDualslopesFrequencydomain2020}. 
\rl{One of the main differences between this technique and others, is its use of only relatively long \glspl{rho} ($\geq\SI{25}{\milli\meter}$) with the hypothesis that data collected at different long \glspl{rho} will feature comparable contributions from superficial (scalp) tissue and different contributions from deeper (brain) tissue.}
\rl{This \gls{DS} technique has been applied primarily to \gls{FD} data,\cite{sassaroliDualslopeMethodEnhanced2019,blaneyPhaseDualslopesFrequencydomain2020,fantiniFrequencyDomainTechniquesCerebral2020} and also has been proposed in \gls{TD}\cite{sawoszMethodImproveDepth2019}.}

The typical \gls{DS} configuration consists of two sources and two detectors which realize symmetric measurements of two slopes of optical data versus \gls{rho}\cite{fantiniTransformationalChangeField2019,blaneyDesignSourceDetector2020}. 
\rl{These slopes are averaged to achieve \gls{DS} measurements that feature a \gls{sen} selective to deeper tissue,\cite{sassaroliDualslopeMethodEnhanced2019,blaneyPhaseDualslopesFrequencydomain2020,fantiniTransformationalChangeField2019} and also supress artifacts from changes in the probe-tissue coupling or from instrumental drifts (inherited from the \gls{SC} method)\cite{hueberNewOpticalProbe1999,blaneyFunctionalBrainMapping2022}.}
\rl{In \gls{FD} \gls{fNIRS}, one measures a \gls{RCom} corresponding to the \gls{fmod} of the source.}
\rl{The slopes of optical data used in \gls{FD} \glspl{DS} are proportional to the differences between measurements at different \glspl{rho} of either the \gls{lnr2I} (also referred to as linearized \gls{I} since $|\acrshort{RCom}|$ is equivalent to \gls{I}) or the phase of the complex reflectance ($\angle\acrshort{RCom}$) (referred to as \gls{phi}).}
\rl{\Gls{DS} also inherits the ability of the \gls{SC} technique to preform calibration-free measurements of absolute optical properties of tissue, namely the \gls{mua} and the \gls{musp}, when the dual slopes of \gls{I} and \gls{phi} are used in combination\cite{hueberNewOpticalProbe1999}.} \par

This work seeks to apply \gls{DS} methods to \gls{DOI} \textit{in-vivo}, bringing with it all of the expected advantages of \gls{DS}. Prior to this work, \gls{DS} \gls{DOI} had been applied to optical phantoms, showing that \gls{DS} \gls{phi} is able to preferentially reconstruct deep perturbations even in the presence of a superficial perturbation\cite{blaneyDualslopeImagingHighly2020}. Extensive work was then done to develop methods to design \gls{DS} \gls{DOI} arrays\cite{blaneyDesignSourceDetector2020}, resulting in the construction of a \gls{DS} array for large area coverage in \gls{fNIRS} \gls{DOI}. 
\q{R2C1a}{\rlB{The methods used in Reference~\citenum{blaneyDesignSourceDetector2020} did not include an analysis on the effect of \glspl{rho} on the \acrfull{sen} \ie{the ratio between a measured \gls{dmua} and a true \gls{dmua} localized within the medium} to top- and bottom- layers, but instead focused on meeting practical requirements based on instrumental limits.}}
\rl{Herein, the novel aspects are the determination of optimal source-detector distances and first applications of a \gls{DS} \gls{DOI} array for \gls{DS} mapping of cerebral hemodynamics \textit{in-vivo}.} 
\q{R2C1b}{\rlB{The results presented here allow for the investigation of \textit{in-vivo} spatial maps of \gls{DS} \gls{I} and \gls{phi}, as compared to previously reported single-location \gls{DS} measurements\cite{blaneyPhaseDualslopesFrequencydomain2020,phamSensitivityFrequencydomainOptical2021a}, and show the applicability of this novel \gls{DS} array to imaging the human brain.}}
\par

\rl{In this manuscript, three experiments investigating \gls{DS} for \gls{DOI} are presented.}
First, an \textit{in-silico}, theoretical investigation of the \glspl{rho} in a \gls{DS} set using an analytical solution to the diffusion equation for two-layer media\cite{blaneyBroadbandDiffuseOptical2022,liemertLightDiffusionTurbid2010}. This experiment is an extension of the work in Reference~\citenum{blaneyDesignSourceDetector2020} with the goal to further examine the choices made in the \gls{DS} array design, and a special emphasis on the optimal \glspl{rho} for \gls{DS} measurements. The second and third experiments are the first \textit{in-vivo} demonstrations of \gls{DS} \gls{DOI} on the human brain. The second experiment is a standard visual stimulation protocol\cite{bejmInfluenceContrastreversingFrequency2019} whose primary aim is to compare the functional hemodynamic response recorded in the primary visual cortex using different \gls{DS} and SD \gls{fNIRS} data-types. The third experiment seeks to demonstrate \gls{DS} \gls{DOI} of the human brain during systemic \gls{ABP} oscillations in a standard \gls{CHS} protocol\cite{fantiniDynamicModelTissue2014,blaneyPhaseDualslopesFrequencydomain2020}. This third experiment is the first \gls{CHS} imaging application to be presented. It is noted that the emphasis of this work is on technology development and the demonstration of the novel \gls{DS} \gls{DOI} technique for mapping hemodynamics in the human brain. Therefore, a single subject was investigated, and more detailed studies of the temporal and spatial features of the cerebral hemodynamics measured with \gls{DS} \gls{DOI} are left to future research conducted on multiple subjects. \par

\section{Methods}\label{sec:meth}
\subsection{Experiments}
\subsubsection{\textit{In-Silico} Simulations of Two-Layer \& Homogeneous Media}\label{meth:exp:insil}
\rl{The first part of this work investigates how the \glspl{rho} used in \gls{FD} \gls{NIRS} measurements affect the depth of the \acrfull{sen}.}
The focus is on \gls{DS}\cite{sassaroliDualslopeMethodEnhanced2019,blaneyPhaseDualslopesFrequencydomain2020} measurements which utilize a set of at least two \glspl{rho}\cite{fantiniTransformationalChangeField2019,blaneyDesignSourceDetector2020}. 
\rl{Therefore an examination of how the maximum or mean \gls{rho} \ie{$\acrshort{rho}_{max}$ or $\bar{\acrshort{rho}}$, respectively} in a \gls{DS} set affect the depth distribution of \gls{sen} was done.}
\rl{To this aim, two sets of diffusion theory based \textit{in-silico} simulations for various \glspl{rho} in a linear-symmetric \gls{DS} set\cite{fantiniTransformationalChangeField2019,blaneyDesignSourceDetector2020}.}
\rl{For the first set, $\bar{\acrshort{rho}}$ was held constant at \SI{30}{\milli\meter} and the difference between the \glspl{rho} \ie{$\Delta\acrshort{rho}$} was varied from \SIrange{5}{50}{\milli\meter} (Table~\ref{tab:simRhos}(left)).}
\rl{In the second set of simulations, $\acrshort{rho}_{max}$ was held constant at \SI{35}{\milli\meter} and $\Delta\acrshort{rho}$ was varied from \SIrange{1}{28}{\milli\meter} (Table~\ref{tab:simRhos}(right)).} \par

\begin{table}[tbh]
    \caption{\rl{\Acrlongpl{rho} in the \textit{in-silico} simulations}} 
    \label{tab:simRhos}
    \begin{center}       
        \begin{threeparttable}
            \centering
            \begin{tabular}{r|S[table-format=2.1]S[table-format=2.1]||r|S[table-format=2.1]S[table-format=2.1]} 
                \multicolumn{3}{c||}{\rl{Fix $\bar{\acrshort{rho}}=\SI{30}{\milli\meter}$}} & \multicolumn{3}{c}{\rl{Fix $\acrshort{rho}_{max}=\SI{35}{\milli\meter}$}} \\
                \acrshortpl{rho} (\si{\milli\meter}) & {$\bar{\acrshort{rho}}$ (\si{\milli\meter})} & {$\Delta\acrshort{rho}$ (\si{\milli\meter})} & \acrshortpl{rho} (\si{\milli\meter}) & {$\bar{\acrshort{rho}}$ (\si{\milli\meter})} & {$\Delta\acrshort{rho}$ (\si{\milli\meter})} \\
                \hline
                $\left[27.5, 32.5\right]$ & 30.0 & 5.0 & $\left[7.0, 35.0\right]$ & 21.0 & 28.0 \\
                $\left[25.0, 35.0\right]$ & 30.0 & 10.0 & $\left[10.0, 35.0\right]$ & 22.5 & 25.0 \\
                $\left[22.5, 37.5\right]$ & 30.0 & 15.0 & $\left[13.0, 35.0\right]$ & 24.0 & 22.0 \\
                $\left[20.0, 40.0\right]$ & 30.0 & 20.0 & $\left[16.0, 35.0\right]$ & 25.5 & 19.0 \\
                $\left[17.5, 42.5\right]$ & 30.0 & 25.0 & $\left[19.0, 35.0\right]$ & 27.0 & 16.0 \\
                $\left[15.0, 45.0\right]$ & 30.0 & 30.0 & $\left[22.0, 35.0\right]$ & 28.5 & 13.0 \\
                $\left[12.5, 47.5\right]$ & 30.0 & 35.0 & $\left[25.0, 35.0\right]$ & 30.0 & 10.0 \\
                $\left[10.0, 50.0\right]$ & 30.0 & 40.0 & $\left[28.0, 35.0\right]$ & 31.5 & 7.0 \\
                $\left[7.5, 52.5\right]$ & 30.0 & 45.0 & $\left[31.0, 35.0\right]$ & 33.0 & 4.0 \\
                $\left[5.0, 55.0\right]$ & 30.0 & 50.0 & $\left[34.0, 35.0\right]$ & 34.5 & 1.0 \\
            \end{tabular}
            \begin{tablenotes}
                \item Acronyms: \Acrfullpl{rho}, mean \acrshort{rho} ($\bar{\acrshort{rho}}$), max \acrshort{rho} ($\acrshort{rho}_{max}$), and difference between \acrshortpl{rho} ($\Delta\acrshort{rho}$)
            \end{tablenotes}
        \end{threeparttable}
    \end{center}
\end{table}

\rl{For each set of \glspl{rho}, more than sixteen thousand (\num{16807}) analytical two-layer simulations were conducted with differing \gls{ztop} and absolute optical properties \ie{\gls{mua} and \gls{musp} of each layer}\cite{blaneyBroadbandDiffuseOptical2022,liemertLightDiffusionTurbid2010}.}
To achieve this, the five two-layer parameters were varied through seven values and all combinations simulated. The \glspl{mua} of each layer were varied in the range \SIrange{0.005}{0.015}{\per\milli\meter}, the \glspl{musp} in the range \SIrange{0.5}{1.5}{\per\milli\meter}, and \gls{ztop} in the range \SIrange{5}{15}{\milli\meter}. A representative semi-infinite homogeneous medium was also simulated\cite{blaneyPhaseDualslopesFrequencydomain2020,continiPhotonMigrationTurbid1997}. This medium had a \gls{mua} of \SI{0.010}{\per\milli\meter} and a \gls{musp} of \SI{1.0}{\per\milli\meter}. \par

Every simulation considered an \gls{dmua} of \SI{0.001}{\per\milli\meter} in the bottom-layer, and a \gls{dmua} of \SI{-0.001}{\per\milli\meter} in the top-layer. 
\rl{Then, the methods discussed below in Sections~\ref{meth:anal:abs}-\ref{meth:anal:rel} were used to simulate the measured \gls{dmua} \ie{the effective $\acrshort{dmua}_{meas}$ obtained from the data assuming that the medium is semi-infinite and homogeneous} considering either the \gls{DS} data or the long \gls{SD} data, for both \gls{I} and \gls{phi}.}
Since preferentially deep \gls{sen} is desired for non-invasive brain measurements, a recovered $\acrshort{dmua}_{meas}$ was considered better when it was closer to \SI{0.001}{\per\milli\meter} (the actual bottom layer \gls{dmua}). For the representative homogeneous medium, the \gls{sen} was calculated using diffusion theory according to the methods described in Reference~\citenum{blaneyPhaseDualslopesFrequencydomain2020} and $\acrshort{dmua}_{meas}$ was considered as the average over all possible \gls{ztop} dividing the top- and bottom-layer perturbations. Comparison to a representative homogeneous medium was done to connect these results to previous work and conclusions drawn from the homogeneous case\cite{fantiniTransformationalChangeField2019}. Additionally, it should be noted that the homogeneous case is still relevant to this work since a homogeneous model is at the core of the \gls{DS} recovery of $\acrshort{dmua}_{meas}$ (Sections~\ref{meth:anal:abs}-\ref{meth:anal:rel}) and the \gls{DS} \gls{DOI} methods (Section~\ref{meth:anal:recon}). The results from this \textit{in-silico} experiment are shown in Section~\ref{res:insil}. \par

\subsubsection{\textit{In-Vivo} Brain Measurements}
\paragraph{Equipment and Human Subject}
All \textit{in-vivo} measurements were performed using an \gls{FD} \gls{NIRS} \gls{ISSv2} which utilizes two \glspl{lam} (\SIlist{830;690}{\nano\meter}) and a \gls{fmod} of \SI{140.625}{\mega\hertz}. For this work the \gls{ISSv2} was configured to use \num{16} source pairs (two \glspl{lam}, thus \num{32} \glspl{LD} total) and \num{10} detectors with a collection sampling rate of \SI{4.96}{\hertz}. \par

\rl{One healthy human subject (\SI{28}{\yr} old male) was recruited for two Tufts University \gls{IRB} approved protocols (expounded upon below).}
The first protocol consisted of visual stimulation (experiment repeated three times), and the second involved systemic \gls{ABP} oscillations (experiment repeated four times). The data presented here are representative of the repeated experiments, which generated similar results. It is noted that only one subject was chosen for this work since the goal is not to draw conclusions about the specific spatio-temporal characteristics of the functional or physiological cerebral hemodynamics measured, but rather to demonstrate the design, advantages, and applicability to the human brain of a \gls{DS} array. \par

\begin{figure}[th]
    \begin{center}
        \begin{tabular}{c}
			\includegraphics[width=0.95\linewidth]{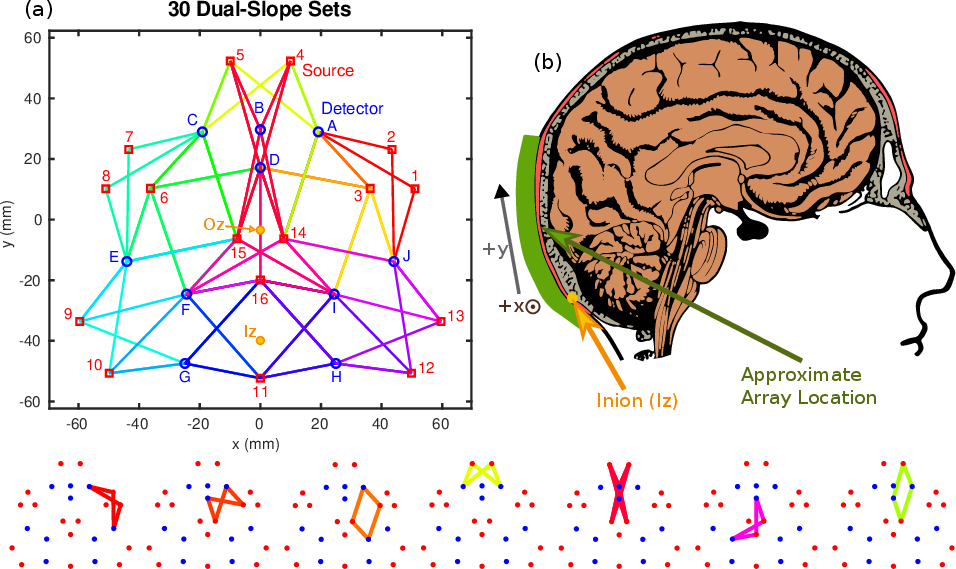}
		\end{tabular}
    \end{center}
    \caption
    {\label{fig:array}
    (a) Schematic of \acrfull{DS} array where lines of differing colors show different \acrshort{DS} sets. \rl{Examples of individual \acrshort{DS} sets are shown along the bottom edge of the figure.} \rl{The approximate \acrfull{Iz} and \acrfull{Oz} locations are shown in yellow.} \rl{The subject's \acrshort{Iz} to \acrfull{Nz} distance was \SI{365}{\milli\meter}.} A full list of \acrshort{DS} sets can be found in Table~\ref{tab:DSinds}. (b) Placement of the \acrshort{DS} array during the \textit{in-vivo} brain measurements showing the upper portion of the array primarily probing the \textit{occipital} lobe.} 
\end{figure}

For both protocols, the \gls{DS} array described in Reference~\citenum{blaneyDesignSourceDetector2020} was placed on the back of the subject's head so that the upper part of the array was over the \textit{occipital} lobe (Figure~\ref{fig:array}(b)). 
\rl{In each experimental repetition, the optical array was placed in approximately the same region as shown in Figure~\ref{fig:array}(b).}
\q{R1C1}{\rlA{The subject's \gls{Iz} to \gls{Nz} distance was approximately \SI{365}{\milli\meter}, and the array locations corresponding to the \gls{Iz} and \gls{Oz} are shown in Figure~\ref{fig:array}(a).}}
This array consisted of \num{57} \gls{SD} sets and \num{30} \gls{DS} sets (Figure~\ref{fig:array}(a)). 
\rl{The array had an overall triangular shape, and covers an area of approximately \SI{120}{\milli\meter} on a side (about \SI{7200}{\square\milli\meter}).}
\rl{All of the \gls{DS} \glspl{rho} pairs were approximately \SIlist{25;37}{\milli\meter}, since this array was designed to homogenize \glspl{rho}\cite{blaneyDesignSourceDetector2020}.} This design choice is based on the simulations described in Sections~\ref{meth:exp:insil}~\&~\ref{res:insil} in this work. \par 

\begin{table}[tbh]
    \caption{Key of \acrlong{DS} set indexes, locations, and optodes (Figure~\ref{fig:array})}
    \label{tab:DSinds}
    \begin{center}       
        \begin{threeparttable}
            \centering
            \begin{tabular}{c||S[table-format=4.0]S[table-format=4.0]|S[table-format=3.0]S[table-format=3.0]S[table-format=3.0]|rrrr|c}
                    \Acrshort{DS} & \multicolumn{2}{c|}{Centroid} & \multicolumn{3}{c|}{Distance from:} & \multicolumn{4}{c|}{Optodes} & Figure~\ref{fig:funTraces} \\
                    Index & {$x$ (\si{\milli\meter})} & {$y$ (\si{\milli\meter})} & {\acrshort{Iz} (\si{\milli\meter})} & {\acrshort{Oz} (\si{\milli\meter})} & {\acrshort{Pz} (\si{\milli\meter})} & & & & & {Panel} \\
                    \hline
                    \rowcolor{DScol1!25}  1 & 39 & 12 & 65 & 42 & 70 & \texttt{ 1} & \texttt{A} & \texttt{J} & \texttt{ 2} &  (k) \\
					\rowcolor{DScol2!25}  2 & 16 & 13 & 55 & 22 & 59 & \texttt{ 3} & \texttt{A} & \texttt{D} & \texttt{14} &  (j) \\
					\rowcolor{DScol3!25}  3 & 22 &  2 & 47 & 23 & 71 & \texttt{ 3} & \texttt{A} & \texttt{I} & \texttt{14} &  (o) \\
					\rowcolor{DScol4!25}  4 & 22 &  2 & 47 & 23 & 71 & \texttt{ 3} & \texttt{D} & \texttt{J} & \texttt{14} &  (p) \\
					\rowcolor{DScol5!25}  5 & 28 & -9 & 42 & 29 & 83 & \texttt{ 3} & \texttt{I} & \texttt{J} & \texttt{14} &  (q) \\
					\rowcolor{DScol6!25}  6 &  0 & 41 & 81 & 44 & 29 & \texttt{ 4} & \texttt{A} & \texttt{C} & \texttt{ 5} &  (a) \\
					\rowcolor{DScol7!25}  7 &  9 & 23 & 64 & 28 & 47 & \texttt{ 4} & \texttt{A} & \texttt{D} & \texttt{14} &  (d) \\
					\rowcolor{DScol8!25}  8 & -9 & 23 & 64 & 28 & 47 & \texttt{ 5} & \texttt{C} & \texttt{D} & \texttt{15} &  (b) \\
					\rowcolor{DScol9!25}  9 & -16 & 13 & 55 & 22 & 59 & \texttt{ 6} & \texttt{C} & \texttt{D} & \texttt{15} &  (f) \\
					\rowcolor{DScol10!25} 10 & -22 &  2 & 47 & 23 & 71 & \texttt{ 6} & \texttt{C} & \texttt{F} & \texttt{15} &  (n) \\
					\rowcolor{DScol11!25} 11 & -22 &  2 & 47 & 23 & 71 & \texttt{ 6} & \texttt{D} & \texttt{E} & \texttt{15} &  (m) \\
					\rowcolor{DScol12!25} 12 & -28 & -9 & 42 & 29 & 83 & \texttt{ 6} & \texttt{E} & \texttt{F} & \texttt{15} &  (l) \\
					\rowcolor{DScol13!25} 13 & -39 & 12 & 65 & 42 & 70 & \texttt{ 7} & \texttt{C} & \texttt{E} & \texttt{ 8} &  (e) \\
					\rowcolor{DScol14!25} 14 & -45 & -36 & 45 & 55 & 115 & \texttt{ 9} & \texttt{E} & \texttt{G} & \texttt{10} &  (w) \\
					\rowcolor{DScol15!25} 15 & -34 & -20 & 40 & 38 & 95 & \texttt{ 9} & \texttt{E} & \texttt{F} & \texttt{15} &  (r) \\
					\rowcolor{DScol16!25} 16 & -25 & -36 & 25 & 41 & 108 & \texttt{10} & \texttt{F} & \texttt{G} & \texttt{16} &  (z) \\
					\rowcolor{DScol17!25} 17 & -12 & -36 & 13 & 35 & 106 & \texttt{11} & \texttt{F} & \texttt{G} & \texttt{16} & (aa) \\
					\rowcolor{DScol18!25} 18 &  0 & -36 &  4 & 33 & 106 & \texttt{11} & \texttt{F} & \texttt{H} & \texttt{16} &  (x) \\
					\rowcolor{DScol19!25} 19 & -0 & -36 &  4 & 33 & 106 & \texttt{11} & \texttt{G} & \texttt{I} & \texttt{16} & (ab) \\
					\rowcolor{DScol20!25} 20 & 12 & -36 & 13 & 35 & 106 & \texttt{11} & \texttt{H} & \texttt{I} & \texttt{16} & (ac) \\
					\rowcolor{DScol21!25} 21 & 25 & -36 & 25 & 41 & 108 & \texttt{12} & \texttt{H} & \texttt{I} & \texttt{16} & (ad) \\
					\rowcolor{DScol22!25} 22 & 45 & -36 & 45 & 55 & 115 & \texttt{12} & \texttt{H} & \texttt{J} & \texttt{13} &  (y) \\
					\rowcolor{DScol23!25} 23 & 34 & -20 & 40 & 38 & 95 & \texttt{13} & \texttt{I} & \texttt{J} & \texttt{14} &  (v) \\
					\rowcolor{DScol24!25} 24 & -4 & -8 & 32 &  6 & 78 & \texttt{14} & \texttt{D} & \texttt{F} & \texttt{16} &  (s) \\
					\rowcolor{DScol25!25} 25 &  0 & -15 & 25 & 12 & 85 & \texttt{14} & \texttt{F} & \texttt{I} & \texttt{15} &  (t) \\
					\rowcolor{DScol26!25} 26 &  4 & -8 & 32 &  6 & 78 & \texttt{15} & \texttt{D} & \texttt{I} & \texttt{16} &  (u) \\
					\rowcolor{DScol27!25} 27 &  4 & 23 & 63 & 27 & 47 & \texttt{ 4} & \texttt{B} & \texttt{D} & \texttt{14} &  (i) \\
					\rowcolor{DScol28!25} 28 &  1 & 23 & 63 & 27 & 46 & \texttt{ 4} & \texttt{B} & \texttt{D} & \texttt{15} &  (c) \\
					\rowcolor{DScol29!25} 29 & -1 & 23 & 63 & 27 & 46 & \texttt{ 5} & \texttt{B} & \texttt{D} & \texttt{14} &  (h) \\
					\rowcolor{DScol30!25} 30 & -4 & 23 & 63 & 27 & 47 & \texttt{ 5} & \texttt{B} & \texttt{D} & \texttt{15} &  (g) \\
            \end{tabular}
            \begin{tablenotes}
                \item Acronyms: \Acrfull{DS}, \acrfull{Iz}, \acrfull{Oz}, and \acrfull{Pz}
                \item Note 1: Row colors correspond to the \acrshort{DS} set colors in Figures~\ref{fig:array},\ref{fig:funTraces},\ref{fig:exTrace},\&\ref{fig:exTrace_noAct}.
                \item Note 2: Optodes refer to ones labeled in Figure~\ref{fig:array}, numbers are sources and letters are detectors.
            \end{tablenotes}
        \end{threeparttable}
    \end{center}
\end{table}

\begin{figure}[bth]
    \begin{center}
        \begin{tabular}{c}
             \includegraphics[width=0.95\linewidth]{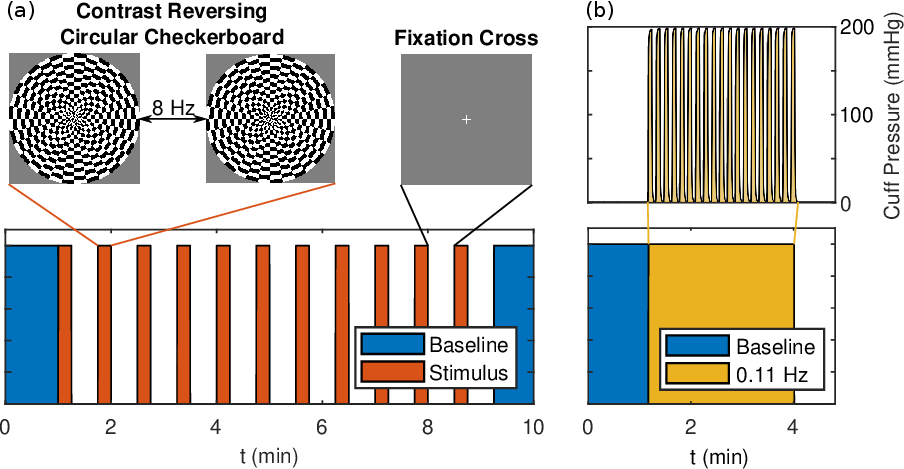}
        \end{tabular}
    \end{center}
    \caption
    {\label{fig:protocols}
    (a) Visual stimulus protocol with \num{11} stimulus periods (\SI{15}{\second} stimulus preceding \SI{30}{\second} rest) using a contrast reversing circular checkerboard, reversing at \SI{8}{\hertz}. (b) Systemic blood pressure oscillations protocol where oscillations at \SI{0.11}{\hertz} lasted for \SI{3}{\min}. The pneumatic thigh \acrshort{cuff} was set to \SI{200}{\milli\meter\Hg} during the cyclic occlusions.}
\end{figure}

\paragraph{Visual Stimulation}
The first \textit{in-vivo} experiment consisted of a visual stimulation protocol. 
\rl{This protocol included an initial baseline and a final recovery baseline, \SI{1}{\min} each, from which absolute optical properties \ie{\gls{mua} and \gls{musp}} were obtained (Section~\ref{meth:anal:abs}).}
\rl{The functional activation portion of the protocol consisted of \num{11} stimulation-rest blocks, where the stimulation lasted \SI{15}{\second} and rest lasted \SI{30}{\second} (Figure~\ref{fig:protocols}(a)).}
\rl{Visual stimulation consisted of a contrast reversing circular checkerboard (\diameter\SI{0.65}{\meter}) which reversed at a frequecy of \SI{8}{\hertz}\cite{bejmInfluenceContrastreversingFrequency2019} and was presented in front of the subject at a distance of \SI{1.8}{\meter}.} Results from this protocol are found in Section~\ref{res:fun}.

\paragraph{Systemic Blood Pressure Oscillations}
The second \textit{in-vivo} experiment consisted of a systemic \gls{ABP} oscillation protocol\cite{blaneyPhaseDualslopesFrequencydomain2020}. 
\q{R1C2}{\rlA{Systemic \gls{ABP} oscillations were induced at a \gls{f} of \SI{0.11}{\hertz} using a \gls{cuff} secured on the upper portion of each of the subject's thighs. The dimensions of each \gls{cuff} were $\SI{180}{\milli\meter} \times \SI{1080}{\milli\meter}$ when laid flat. The \glspl{cuff} were placed so that they were centered on the thighs and secured not to shift during the inflation and deflation procedures.}} 
The amplitude of the \gls{cuff} pressure oscillations was set to \SI{200}{\milli\meter\Hg} (Figure~\ref{fig:protocols}(b)). Continuous \gls{ABP} measurements were taken throughout the experiment using a \gls{CNAP}.
\rl{The \gls{CNAP} achieved these \gls{ABP} measurements using finger plethysmography.}
\rl{This experimental protocol started with a \SI{1}{\min} baseline that was used to find baseline tissue optical properties \ie{\gls{mua} and \gls{musp}; Section~\ref{meth:anal:abs}}.}
\rl{Following the initial baseline, the oscillation sequence began and lasted \SI{3}{\min}, leading to \num{19} oscillation periods at the set frequency of \SI{0.11}{\hertz}.}
The results from this experiment are presented in Section~\ref{res:osc}. \par

\subsection{Analysis}\label{meth:anal}
\subsubsection{Recovery of Absolute Optical Properties}\label{meth:anal:abs}
\rl{Tissue absolute \gls{mua} and \gls{musp} were calculated for each \gls{DS} set, both for the \textit{in-silico} simulations and throughout the \gls{DS} array \textit{in-vivo} measurements.}
This was achieved by using the \gls{DS} set in \gls{SC} \gls{FD}-\gls{NIRS} mode. 
\rl{To convert the \gls{FD} slopes to \gls{mua} and \gls{musp}, an iterative method\cite{blaneyDualslopeDiffuseReflectance2021} based on a semi-infinite homogeneous medium and extrapolated boundary conditions\cite{continiPhotonMigrationTurbid1997} was used.}
\rl{Briefly, this method uses the \gls{RCom} versus \gls{rho}, an initial guess of the \gls{mueffCom} using assumptions of linearity\cite{fantiniNoninvasiveOpticalMonitoring1999}, and finds \gls{mua} and \gls{musp} by iteratively solving the analytical equation for \gls{RCom} in a semi-infinite homogeneous medium\cite{blaneyDualslopeDiffuseReflectance2021}.} 
\rl{The iteratively recovered \gls{mueffCom} was then converted to \gls{mua} and \gls{musp} for each \gls{DS} set.} \par

\subsubsection{Measuring Changes in Hemoglobin Concentration}\label{meth:anal:rel}
Dynamic changes in \gls{I} or \gls{phi} for \gls{SD} or \gls{DS} were translated into \gls{dmua} using methods reported in Reference~\citenum{blaneyPhaseDualslopesFrequencydomain2020}. 
\rl{For \gls{SD}, \gls{dmua} was calculated using the \gls{DPF} obtained using the absolute \gls{mua} and \gls{musp} calculated as described in Section~\ref{meth:anal:abs}.}
In the case of \gls{DS}, \gls{dmua} was calculated using the \gls{DSF} obtained from said \gls{mua} and \gls{musp}. 
\rl{These measured \glspl{dmua} at two \glspl{lam} were converted to \gls{dO} and \gls{dD} using known hemoglobin extinction coefficients and Beer's law\cite{prahlTabulatedMolarExtinction1998}.} \par

\subsubsection{Phasor Analysis}\label{meth:anal:phs}
The systemic \gls{ABP} oscillations experiment presented in Section~\ref{res:osc} required transfer function analysis for interpretation. 
\rl{This was performed for each \gls{SD} and \gls{DS} set and each data-type (\gls{I} or \gls{phi}) independently.}
\rl{The analysis was done to retrieve a \gls{OAvec}, and a \gls{DAvec} at the induced frequency of \SI{0.11}{\hertz}.}
\rl{These vectors represented the amplitude ratio \ie{modulus} and the phase difference \ie{argument} of the two signals considered.}
\rl{To achieve this, the \gls{CWT} (\texttt{cwt} function in \gls{MATLAB}), based on a complex Morlet mother wavelet, was taken of the temporal \ie{\gls{t}} signals \gls{dO}, \gls{dD}, and \gls{dABP}.}
The wavelet coefficients were interpreted as phasor maps of the \gls{Oph}, \gls{Dph}, and \gls{ABPph} over \gls{t} and \gls{f}. Then the quotient from division of the corresponding phasor maps created the transfer functions \gls{OAvec} and \gls{DAvec} also over \gls{t} and \gls{f}. \par

\rl{To identify which \gls{t} and \gls{f} regions to use in further analysis, the wavelet \gls{OAcoh} and the \gls{DAcoh} were calculated using a modified version of the \gls{MATLAB} \texttt{wcoherence} function, which removes smoothing in \gls{f}.}
A \gls{coh} threshold generated from the \num{95}\textsuperscript{th} percentile \ie{$\alpha=\num{0.05}$} of \gls{coh} between random surrogate data\cite{blaneyAlgorithmDeterminationThresholds2020} was used to mask both \gls{OAcoh} and \gls{DAcoh}  maps so that only \glspl{t} and \glspl{f} with significant coherence between the considered signals contained Boolean \emph{true}. Next, a logical \emph{and} was taken between both threshold-ed \gls{OAcoh} and \gls{DAcoh} Boolean maps, so that only \glspl{t} and \glspl{f} in which both \gls{dO} and \gls{dD} were coherent with \gls{dABP} retained \emph{true} Boolean values\cite{khaksariDepthDependenceCoherent2018,phamNoninvasiveOptical2021}. 
\rl{This Boolean map of significant \gls{coh} was then used to mask the \gls{OAvec} and \gls{DAvec} transfer function maps, allowing only transfer function relationships of significant coherence to be considered in the analysis.} \par

\rl{To select only \glspl{f} around the induced frequency of \SI{0.11}{\hertz}, the bandwidth of a test sinusoid extending the duration of the protocol was found to be \SI{0.02}{\hertz} using the \gls{FWHM} of the \gls{CWT} amplitude\cite{phamNoninvasiveOptical2021}.}
Finally, the significant \gls{coh} masked transfer functions, \gls{OAvec} and \gls{DAvec}, were averaged within this \gls{f} band and during the induced oscillation \gls{t} window (Figure~\ref{fig:protocols}(b)). 
\rl{Therefore, the results reflected measured hemodynamics that featured significant \gls{coh} for both \gls{OAcoh} and \gls{DAcoh} at the frequency induced (\SI{0.11}{\hertz}).} \par

\subsubsection{Image Reconstruction}\label{meth:anal:recon}
\paragraph{General Imaging Methods}
For image reconstruction the \gls{MPi} was implemented with Tikhonov regularization (scaling parameter $a=1$)\cite{blaneyDesignSourceDetector2020, blaneyDualslopeImagingHighly2020}. 
\rl{Reconstruction was conducted on the \gls{SD} \gls{I}, \gls{SD} \gls{phi}, \gls{DS} \gls{I}, and \gls{DS} \gls{phi} data separately, creating a different image for each data-type and allowing comparisons between them.}
\num{57} \gls{SD} and \num{30} \gls{DS} sets existed in the array (Figure~\ref{fig:array}); however only sets which passed data quality requirements were considered for reconstruction (Section~\ref{meth:anal:qual}). The \gls{senMat} (which was inverted with \gls{MPi}) was generated considering a semi-infinite homogeneous medium\cite{blaneyDesignSourceDetector2020} and the local \gls{DS} measured optical properties (also used for calculation of \gls{DPF} or \gls{DSF}, Sections~\ref{meth:anal:abs}-\ref{meth:anal:rel}). The medium was voxelized using two layers of pillars (voxels long in $z$ \ie{depth}) with a lateral pitch of \SI{1}{\milli\meter} (along $x$ and $y$), and an axial size (along $z$) of \SI{5}{\milli\meter} for the top-layer and \SI{25}{\milli\meter} for the bottom-layer. The images reported here represent reconstructed values of the bottom-layer voxels in the $x-y$ plane. 
\rl{This method for voxelizing the medium was used before for \gls{DS} imaging in References~\citenum{blaneyDesignSourceDetector2020}~\&~\citenum{blaneyDualslopeImagingHighly2020}.} \par

\paragraph{Visual Stimulation Imaging}
\rl{For the visual stimulation protocol, image reconstruction was conducted on the \gls{dmua} for each time-point (for each \gls{lam}), resulting in an image stack of \gls{dmua}.}
Then, Beer's law was used to create image stacks of \gls{dO} and \gls{dD} as discussed in Section~\ref{meth:anal:rel}. These maps were then temporally folding averaged over the \num{11} stimulus and rest periods (Figure~\ref{fig:protocols}). Two \SI{10}{\second} temporal windows of the image stack were selected to represent the stimulus and rest, respectively. The stimulus window ended \SI{1}{\second} before the end of the \SI{15}{\second} stimulus, and the rest window was centered in the \SI{30}{\second} rest period. Considering these temporal windows of image stacks, a \textit{t}-test ($\alpha=\num{0.05}$) was conducted for every pixel. For \gls{dO} the alternate hypothesis was that stimulus \gls{dO} was greater then rest \gls{dO}, while for \gls{dD} it was that stimulus \gls{dD} was less than rest \gls{dD}. A significant activation Boolean spatial mask was made by only considering pixels where both \gls{dO} significantly increased and \gls{dD} significantly decreased during stimulus compared to rest. In addition to the Boolean mask, an activation amplitude image was also created. For this, the image stacks of \gls{dO} and \gls{dD} were subtracted resulting in a image stack of \gls{dEth} (a surrogate measurement of \gls{BF})\cite{tsujiInfraredSpectroscopyDetects1998,phamQuantitativeMeasurementsCerebral2019}. The average \gls{dEth} was found for both the stimulus and rest \SI{10}{\second} windows, and the activation amplitude was taken to be the difference between the two \ie{$\Delta \text{\textit{\DH}}_{stim} - \Delta \text{\textit{\DH}}_{rest}=(\acrshort{dO}_{stim}-\acrshort{dO}_{rest})+(\acrshort{dD}_{rest}-\acrshort{dD}_{stim})$}. This \gls{dEth} amplitude map was masked by the Boolean mask of significant activation found via \textit{t}-test to result in the activation images presented (Section~\ref{res:fun}). \par

\paragraph{Systemic \gls{ABP} Oscillations Phasor Imaging}
The systemic \gls{ABP} oscillation protocol required a different workflow to result in reconstructed images. 
\rl{The images sought in this case were maps of the amplitude and phase of the \gls{DOvec}, and the \gls{TAvec} at \SI{0.11}{\hertz}.}
The methods in Section~\ref{meth:anal:phs} output \gls{OAvec} and \gls{DAvec} for each measurement set in the array. Using Beer's law\cite{prahlTabulatedMolarExtinction1998}, these were converted to the \gls{muaAvec} at each \gls{lam}. From here, \gls{MPi} was applied to the same \gls{senMat} as above and image reconstruction was conducted on the complex numbers representing \gls{muaAvec}. These spatial maps of \gls{muaAvec} at the two \gls{lam} were then converted to maps of \gls{OAvec} and \gls{DAvec}, again using Beer's law\cite{prahlTabulatedMolarExtinction1998}. 
\rl{Finally, maps of \gls{DOvec} were created using the ratio of \gls{OAvec} and \gls{DAvec}, and maps of \gls{TAvec} using their sum.}
These maps of complex numbers were then smoothed using a Gaussian filter with characteristic length equal to the average array resolution\cite{blaneyDesignSourceDetector2020} to remove artifacts created by applying \gls{MPi} to complex numbers. The amplitude and phase of these maps of \gls{DOvec} and \gls{TAvec} are visualized and presented herein (Section~\ref{res:osc}). \par

\subsubsection{Data Quality Evaluation}\label{meth:anal:qual}
\rl{To ensure that only sufficiently good-quality data were used for further analysis and image reconstruction, each data set was tested in terms of noise, coherence, signal amplitude, or voxel sensitivities.}
Bad sets were eliminated so they were not considered in analysis and their sensitivity region not included in \gls{senMat} \ie{the region under a bad set did not contain voxels used in image reconstruction}. For both the visual stimulation and systemic \gls{ABP} oscillations, a threshold on the noise was applied. This threshold was evaluated by first high-pass filtering to \SI{1.7}{\hertz} \ie{above heart rate} to eliminate power from physiological oscillations. Then the average of the sliding windowed standard deviation (window size of \SI{10}{\second}) was taken as the noise amplitude (corrected for power lost at low \gls{f} from the filter, assuming white noise). 
\q{R1C6a}{\rlA{Channels with higher noise amplitude than \SI{1}{\micro\Molar} in \gls{dT} were considered bad and excluded from further analysis as noise of this amplitude would dominate over responses associated with functional of physiological cerebral hemodynamics.}} \par

\rl{For the \gls{ABP} oscillations data, further quality evaluation was conducted beyond the wavelet \gls{coh} analysis described in Section~\ref{meth:anal:phs}\cite{blaneyAlgorithmDeterminationThresholds2020}.}
\rl{Any voxels with \gls{sen} below the \num{1}\textsuperscript{st} percentile of all \gls{sen} in \gls{senMat} were ignored, as well as any measurement pairs that measured less than \SI{0.001}{\micro\Molar\per\milli\meter\per\Hg} in amplitude.}
\rl{The reasoning for the former being that voxels with small \gls{sen} will create large artifacts in image reconstruction (only partially addressed by Tikhonov regularization), and the reason for the latter being that one cannot claim that such a small amplitude transfer function vector was measurable considering the noise in the system.}
\q{R1C6b}{\rlA{In fact, an amplitude below \SI{0.001}{\micro\Molar\per\milli\meter\per\Hg} would correspond to an immeasurably small oscillation in cerebral hemodynamics on the order of \SI{0.01}{\micro\Molar} considering a typical \gls{ABP} oscillation amplitude on the order of \SI{10}{\milli\meter\per\Hg}.}}
\par

\section{Results}\label{sec:res}
\subsection{\textit{In-Silico} Investigation of Optimal Source-Detector Distances}\label{res:insil}
\begin{figure}[bth]
    \begin{center}
        \begin{tabular}{c}
            \includegraphics[width=0.95\linewidth]{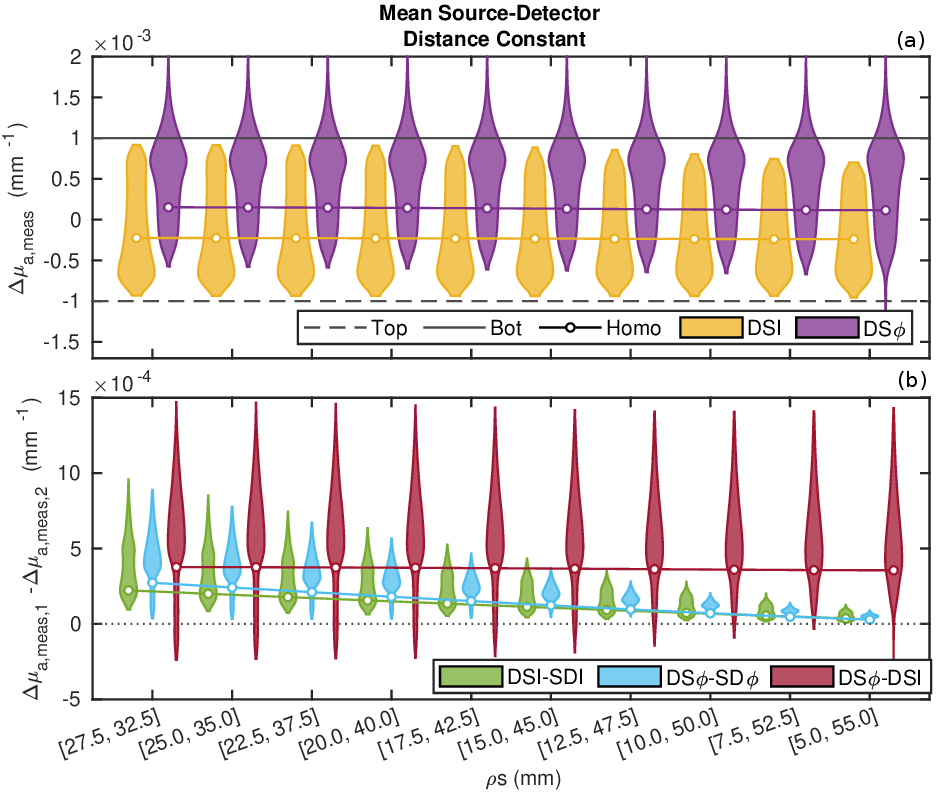}
        \end{tabular}
    \end{center}
    \caption
    {\label{fig:violins_meanConst}
    Simulated \acrfull{dmua} obtained from various media. Specifics of the \acrfullpl{rho} including their means, maxima, and differences are shown in Table~\ref{tab:simRhos}. \num{16807} simulations of different two-layer media were conducted (for each set of \acrshort{rho}), plotted as violins. Also, a representative medium with homogeneous (homo) absolute optical properties was simulated and the average measured \acrshort{dmua} (averaged over all possible layer thicknesses) recovered. In all cases the actual \acrshort{dmua} was \SI{-0.001}{\per\milli\meter} in the top layer and \SI{0.001}{\per\milli\meter} in the bottom layer. \textbf{The two \glspl{rho} in a \acrfull{DS} set were varied such that their mean \acrshort{rho} was constant and their difference increased.} (a) Violin plots (from all \num{16807} two-layer simulations) and lines with points (from example homogeneous simulation) showing the measured \acrshort{dmua} from \acrshort{DS} \acrfull{I} and \acrshort{DS} \acrfull{phi}. (b) Violin plots and lines (all \num{16807} two-layer and example homogeneous, respectively) showing differences between two measurement types, either \acrshort{DS}\acrshort{I} and long \acrfull{SD} \acrshort{I}, \acrshort{DS}\acrshort{phi} and long \acrshort{SD}\acrshort{phi}, or \acrshort{DS}\acrshort{phi} and \acrshort{DS}\acrshort{I}.}
\end{figure}

\begin{figure}[bth]
    \begin{center}
        \begin{tabular}{c}
            \includegraphics[width=0.95\linewidth]{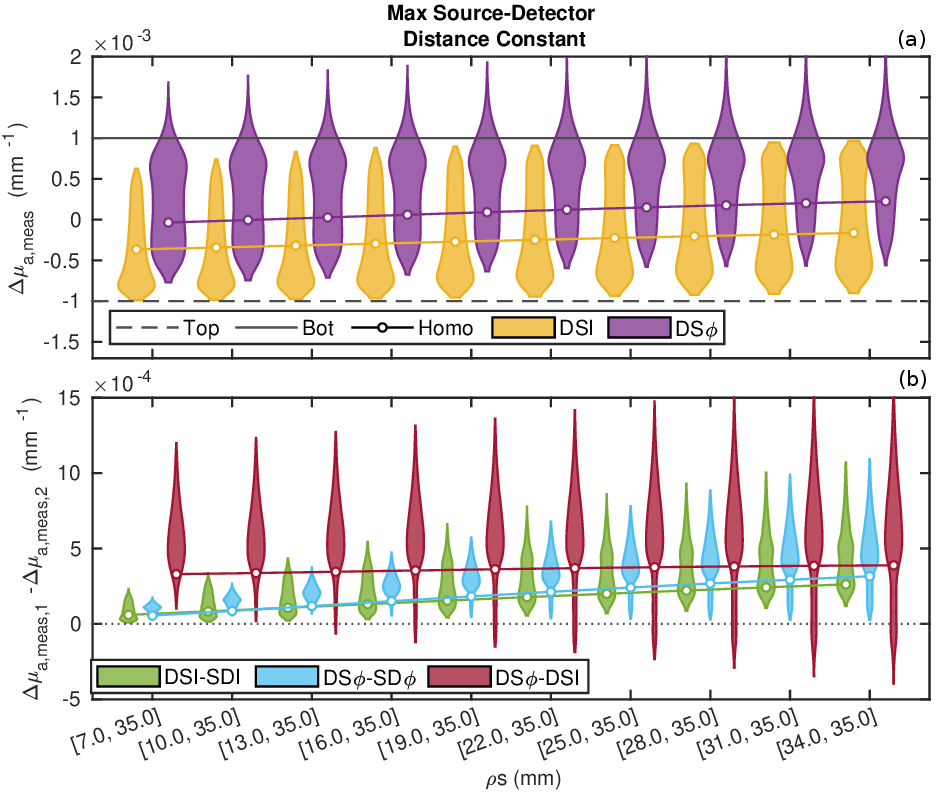}
        \end{tabular}
    \end{center}
    \caption
    {\label{fig:violins_maxConst}
    Simulated \acrfull{dmua} obtained from various media. Specifics of the \acrfullpl{rho} including their means, maxima, and differences are shown in Table~\ref{tab:simRhos}. \num{16807} simulations of different two-layer media were conducted (for each set of \acrshort{rho}), plotted as violins. Also, a representative medium with homogeneous (homo) absolute optical properties was simulated and the average measured \acrshort{dmua} (averaged over all possible layer thicknesses) recovered. In all cases the actual \acrshort{dmua} was \SI{-0.001}{\per\milli\meter} in the top layer and \SI{0.001}{\per\milli\meter} in the bottom layer. \textbf{The two \glspl{rho} in a \acrfull{DS} set were varied such that their maxium \acrshort{rho} was constant and their difference decreased.} (a) Violin plots (from all \num{16807} two-layer simulations) and lines with points (from example homogeneous simulation) showing the measured \acrshort{dmua} from \acrshort{DS} \acrfull{I} and \acrshort{DS} \acrfull{phi}. (b) Violin plots and lines (all \num{16807} two-layer and example homogeneous, respectively) showing differences between two measurement types, either \acrshort{DS}\acrshort{I} and long \acrfull{SD} \acrshort{I}, \acrshort{DS}\acrshort{phi} and long \acrshort{SD}\acrshort{phi}, or \acrshort{DS}\acrshort{phi} and \acrshort{DS}\acrshort{I}.}
\end{figure}

Figures~\ref{fig:violins_meanConst}-\ref{fig:violins_maxConst} report the \gls{dmua} obtained from data computed with diffusion theory for the various conditions described in Section~\ref{meth:exp:insil}. The subplots in Figure~\ref{fig:violins_meanConst} show the results for a constant mean \acrlong{rho} ($\bar{\acrshort{rho}}=\SI{30}{\milli\meter}$), and the subplots in Figure~\ref{fig:violins_maxConst} show the results for a constant maximum \acrlong{rho} ($\acrshort{rho}_{max}=\SI{35}{\milli\meter}$) (Section~\ref{meth:exp:insil} and Table~\ref{tab:simRhos}). The data in Figures~\ref{fig:violins_meanConst}-\ref{fig:violins_maxConst} are reported with violin plots, which show the probability density (represented by the violin thickness) corresponding to each \gls{dmua} value along the vertical axis.
\q{R2C2a}{\rlB{For these simulations there are two true \glspl{dmua}, one of the bottom- and one of the top- layer. In general, the goal is to measure a value of \gls{dmua} that is close to that of the bottom-layer.}}
\q{R2C2b}{\rlB{The reader is reminded that the homogeneous medium considered here is homogeneous in absolute optical properties but heterogeneous in \gls{dmua}. Understanding the recovered value for this case is quite straightforward as it is a weighted average of the \glspl{sen} presented previously.\cite{sassaroliDualslopeMethodEnhanced2019,fantiniTransformationalChangeField2019,blaneyPhaseDualslopesFrequencydomain2020} This simpler interpretation is the motivation for including this case in the simulations, and to allow one to connect the new results to previous work.}}
\par

\rl{First, in Figures~\ref{fig:violins_meanConst}(a)\&\ref{fig:violins_maxConst}(a), which report the \gls{dmua} recovered from \gls{DS} \gls{I} or \gls{DS} \gls{phi}, one can see that in the homogeneous medium (shown by the solid line with a circle) and two-layer simulations (violin plots) there is no notable difference amongst the \gls{rho} sets where $\bar{\acrshort{rho}}$ is constant.}
\rl{On the other hand, when keeping $\acrshort{rho}_{max}$ constant, as the two \gls{SD} that comprise the \gls{DS} set become closer to each other \ie{$\Delta\acrshort{rho}$ becomes smaller}, most recovered \glspl{dmua} become closer to that of the bottom layer.}
\rl{This trend is apparent in both the two-layer media (violin plots), as the mode of the \gls{dmua} distribution in two-layered media (violin plots) increases towards the actual bottom layer \gls{dmua}, and in the homogeneous medium as the solid line also approaches the actual bottom layer \gls{dmua}.}
From this, it is apparent that what mostly affects the sensitivity depth of \gls{DS} is $\bar{\acrshort{rho}}$ and not $\acrshort{rho}_{max}$. The mode (the visually easy part of the violin plot to identify) of the \gls{dmua} distribution in two-layered media (violin plots) obtained with \gls{DS} \gls{phi} is always closer to the actual bottom layer \gls{dmua} compared to the mode obtained with \gls{DS} \gls{I} and compared to the values obtained with \gls{DS} \gls{phi} or \gls{DS} \gls{I} in the representative homogeneous medium. It is also worth noting that the mode of the \gls{dmua} distribution in two-layered media obtained with \gls{DS} \gls{I} is always closer to the actual top layer \gls{dmua} compared to the \gls{dmua} recovered in the homogeneous medium. 
\rl{One can conclude that, in general \ie{for the majority of simulations}, \gls{DS} \gls{phi} is more sensitive to the bottom layer compared to \gls{DS} \gls{I}, and that the \gls{DS} \gls{phi} more closely retrieves the \glspl{dmua} that occur deeper in a two-layered medium than in a homogeneous medium.} \par

\rl{The difference between data-types can be evaluated by examining Figures~\ref{fig:violins_meanConst}(b)\&\ref{fig:violins_maxConst}(b).}
\rl{Note that here only the longer \gls{SD} \ie{\gls{SD} data that feature the deepest sensitivity} is considered in the differences.}
\rl{In the case of \gls{DS} \gls{I} minus \gls{SD} \gls{I} (green) and \gls{DS} \gls{phi} minus \gls{SD} \gls{phi} (blue), in both sets of simulations (two-layered media and homogeneous medium), the difference between data-types is positive and moves toward zero as $\Delta\acrshort{rho}$ increases.}
In Figure~\ref{fig:violins_meanConst}(b), where the \gls{DS} sensitivity depth is about constant because of the constant $\bar{\acrshort{rho}}$, this is due to the increase in depth sensitivity of \gls{SD} data at longer \glspl{rho}. In Figure~\ref{fig:violins_maxConst}(b), where the \gls{SD} sensitivity depth is constant because of the constant $\acrshort{rho}_{max}$, this is due to the decrease in depth sensitivity of \gls{DS} data as $\Delta\acrshort{rho}$ becomes larger. It is important to note, however, that \gls{DS} data (both \gls{I} and \gls{phi}) always result in a greater \acrshort{dmua} than the corresponding \gls{SD} data, indicating a stronger sensitivity to the bottom layer achieved by \gls{DS} versus \gls{SD} data. Now focusing on \gls{DS} \gls{phi} and \gls{DS} \gls{I}, the difference between the associated values of \acrshort{dmua} (red) across different sets of \glspl{rho} is almost constant  (Figure~\ref{fig:violins_meanConst}(b)\&\ref{fig:violins_maxConst}(b)). This indicates that variations in neither $\bar{\acrshort{rho}}$ nor $\acrshort{rho}_{max}$ significantly affect the relationship between the \glspl{dmua} measured by the two \glspl{DS} types. The caveat being that Figure~\ref{fig:violins_maxConst}(b) shows a slightly greater improvement in the depth sensitivity of \gls{DS} \gls{phi} compared to \gls{DS} \gls{I} as $\Delta\acrshort{rho}$ decreases. However, a clear result is that, as $\Delta\acrshort{rho}$ becomes smaller, the variance of the difference between \glspl{dmua} from pairs of data-types increases. 
\rl{Furthermore, there are special cases (expounded upon in Section~\ref{sec:dis}) in which \gls{DS} \gls{I} achieves better sensitivity to the bottom layer compared to \gls{DS} \gls{phi} (as indicated by the portions of the violin plots below zero in Figure~\ref{fig:violins_meanConst}(b)\&\ref{fig:violins_maxConst}(b)), but in general this is not the case.} \par

\subsection{\textit{In-Vivo} Visual Stimulation}\label{res:fun}
\begin{figure}[bth]
    \begin{center}
        \begin{tabular}{c}
			\includegraphics[width=0.95\linewidth]{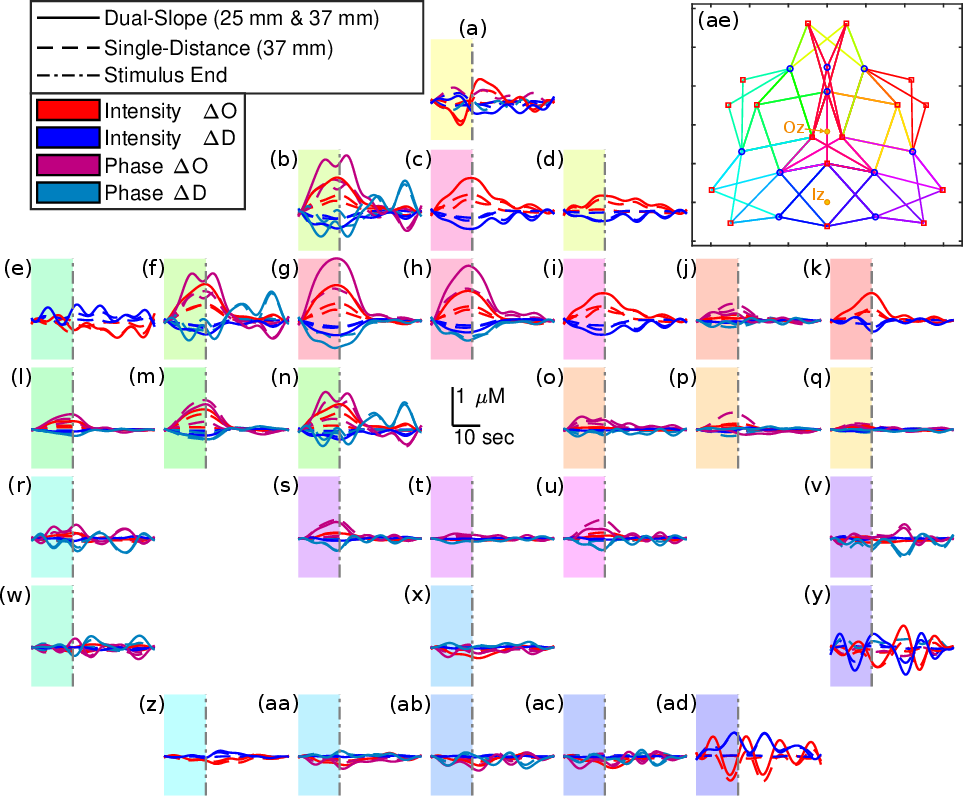}
		\end{tabular}
    \end{center}
    \caption
    {\label{fig:funTraces}
    (a)-(ad) \Acrfull{dO} and \acrfull{dD} of \num{11} repetition folding averages from an example \textit{in-vivo} visual stimulation experiment. \rl{Locations of the insets approximately correspond to the location of each measurement set.} Shaded regions indicate the activation period. A key to connect each panel to the specific \acrfull{DS} set can be found in Table~\ref{tab:DSinds}. (ae) Schematic of array from Figure~\ref{fig:array} indicating the approximate \acrfull{Iz} and \acrfull{Oz} location. Colors of the \acrshort{DS} sets in (ae) correspond to the color of the shaded region in (a)-(ad) which also correspond to the row colors in Table~\ref{tab:DSinds}. Note: Traces are low-passed to \SI{0.1}{\hertz}.}
\end{figure}

Figure~\ref{fig:funTraces} shows the activation traces over the entire \gls{DS} array from an example data-set of the repeated visual stimulation experiments. Each subplot represents one \gls{DS} where the plot locations approximately correspond to the set location on the subject's head. The subplots also show the data collected at the two long \gls{SD} (\gls{rho} of \SI{37}{\milli\meter}) pairs within each \gls{DS} set. These traces are the result of low-pass filtering the data at \SI{0.1}{\hertz} then folding averaging across the \num{11} stimulation periods. The first \SI{15}{\second} of the trace represent the visual stimulation, whereas the following \SI{30}{\second} represent the rest period (Figure~\ref{fig:protocols}(a)). For Figure~\ref{fig:funTraces}, the characteristic functional activation response (increase in \gls{dO} and decrease in \gls{dD}) is localized to the upper center of the array. Additionally, for channels associated with activation, the amplitude of the functional hemodynamic response is greatest when measured with \gls{DS} \gls{phi} followed by \gls{DS} \gls{I} or \gls{SD} \gls{phi} then \gls{SD} \gls{I} (noting that \gls{SD} \gls{I} is the typical measurement used by \gls{CW} \gls{fNIRS}). The oscillations such as the ones in Figure~\ref{fig:funTraces}(ad) are attributed to noise and to the cut-off frequency of the low-pass filter, showing that the \SI{1}{\micro\Molar} noise threshold (Section~\ref{meth:anal:qual}) may still allow noisy channels through the analysis. \par

\begin{figure}[bth]
    \begin{center}
        \begin{tabular}{c}
			\includegraphics[width=0.95\linewidth]{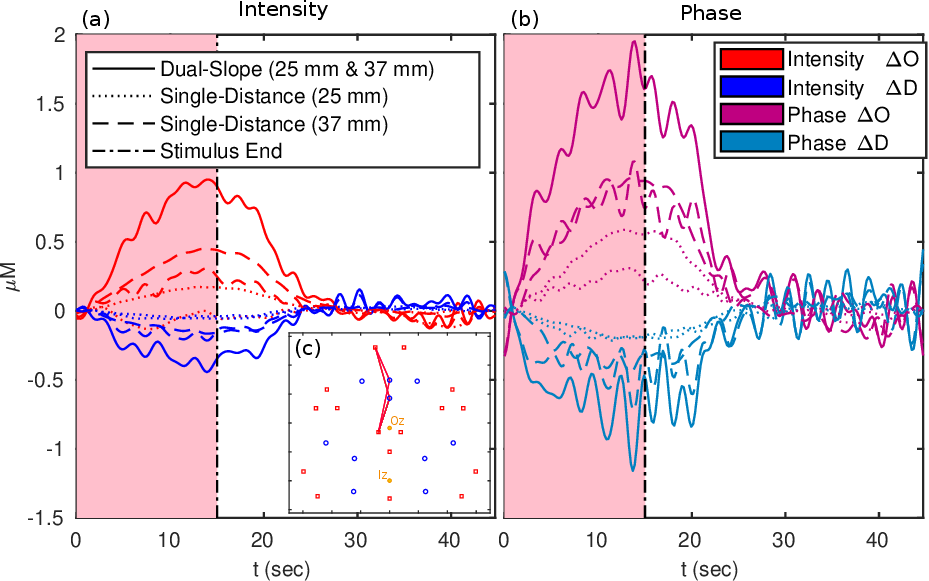}
		\end{tabular}
    \end{center}
    \caption
    {\label{fig:exTrace}
    (a)\&(b) \Acrfull{dO} and \acrfull{dD} of an \num{11} repetition folding average from \acrfull{DS} index 30 (Table~\ref{tab:DSinds}) shown in Figure~\ref{fig:funTraces}(g). Shaded region indicates activation period and color corresponds to the color of the \acrfull{DS} set in Figure~\ref{fig:array}. (c) The location of the set plotted with the approximate \acrfull{Iz} and \acrfull{Oz} location indicated. Note: Traces are low-passed to \SI{0.5}{\hertz}.}
\end{figure}

\begin{figure}[bth]
    \begin{center}
        \begin{tabular}{c}
			\includegraphics[width=0.95\linewidth]{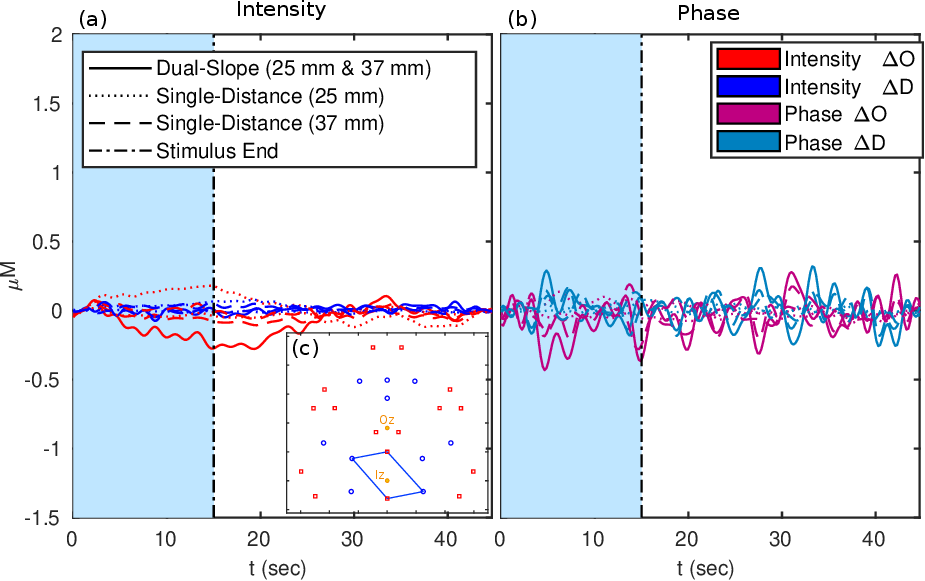}
		\end{tabular}
    \end{center}
    \caption
    {\label{fig:exTrace_noAct}
    (a)\&(b) \Acrfull{dO} and \acrfull{dD} of an \num{11} repetition folding average from \acrfull{DS} index 18 (Table~\ref{tab:DSinds}) shown in Figure~\ref{fig:funTraces}(x). Shaded region indicates activation period and color corresponds to the color of the \acrfull{DS} set in Figure~\ref{fig:array}. (c) The location of the set plotted with the approximate \acrfull{Iz} and \acrfull{Oz} location indicated. Note: Traces are low-passed to \SI{0.5}{\hertz}.}
\end{figure}

\rl{Figures~\ref{fig:exTrace}\&\ref{fig:exTrace_noAct}, show a zoomed-in folding average of the \gls{DS} data set reported in Figures~\ref{fig:funTraces}(g)\&(x), respectively.}
\rl{These traces include all the measurements shown in Figures~\ref{fig:funTraces}(g)\&(x), with the addition of the short \gls{SD} within the \gls{DS} set (\gls{rho} of \SI{25}{\milli\meter}).}
In this case, the traces are low-pass filtered to \SI{0.5}{\hertz} then folding averaged over the \num{11} stimulus periods. 
\q{R2C4}{\rlB{The oscillations in Figures~\ref{fig:exTrace}\&\ref{fig:exTrace_noAct} are due to the noise in the signal (evident in \gls{DS} \gls{phi} due to a higher noise of \gls{phi} data overall) and the cut-off frequency of the filter (\SI{0.5}{\hertz}).}}\par

\q{R1C5}{\rlA{Figure~\ref{fig:exTrace} is an example of a \gls{DS} set with significant activation while Figure~\ref{fig:exTrace_noAct} is an example of a set without significant activation.}}
As with the amplitude relations noted in Figure~\ref{fig:funTraces}, Figure~\ref{fig:exTrace} also shows \gls{DS} \gls{phi} resulting in the hemodynamic response with the largest amplitude; this is followed by \gls{DS} \gls{I} or long \gls{SD} \gls{phi}, then long \gls{SD} \gls{I} or short \gls{SD} \gls{phi}, and then short \gls{SD} \gls{I}. It is worth noting that one short \gls{SD} \gls{I} data does not exhibit an activation signature, and both short \gls{SD} \glspl{I} measured almost no decrease in \gls{dD}  (again, \gls{SD} \gls{I} in the most common data-type used in typical \gls{CW} \gls{fNIRS}). Despite this, both short \gls{SD} \glspl{phi} did measure said activation signature (including a decrease in \gls{dD}) despite the measurements coming from the exact same optodes used to collect \gls{SD} \glspl{I}. \par

\begin{figure}[bth]
    \begin{center}
        \begin{tabular}{c}
			\includegraphics[width=0.95\linewidth]{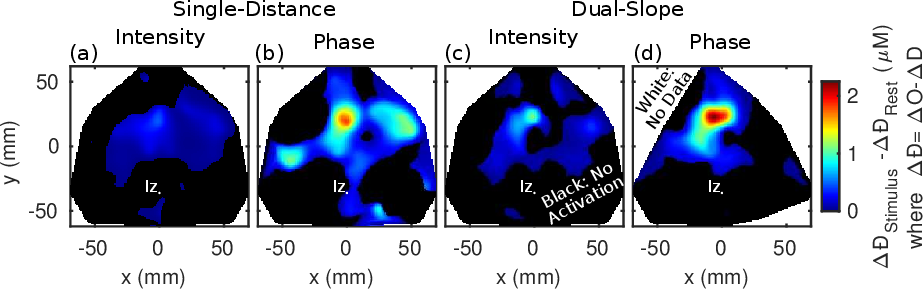}
		\end{tabular}
    \end{center}
    \caption
    {\label{fig:funActMap}
    Maps of activation amplitude during the example \textit{in-vivo} visual stimulation experiment across data-types: (a) \acrfull{SD} \acrfull{I}, (b) \acrshort{SD} \acrfull{phi}, (c) \acrfull{DS} \acrshort{I}, and (d) \acrshort{DS}\acrshort{phi}. The approximate \acrfull{Iz} location is indicated. The activation amplitude is taken to be the difference between \acrfull{dEth} averages from \SI{10}{\second} windows during stimulus and rest. White areas show where there is no data after data quality restrictions were enforced and black areas show where activation was not significant.}
\end{figure}

The final results figure for the visual stimulation protocol is an activation image (Figure~\ref{fig:funActMap}). 
\q{R1C7a}{\rlA{All images were reconstructed using the \gls{MPi} as described in Section~\ref{meth:anal:recon}, with \num{57} \gls{SD} pairs or \num{30} \gls{DS} sets used for their respective reconstructions.}}
\q{R2C5a}{\rlB{Following the methods in Section~\ref{meth:anal:recon}, the black regions of the image are areas with no significant activation, and white areas indicate locations in which no data were present (because they were either not measured or eliminated as described in Section~\ref{meth:anal:qual}). Significant activation was based on requiring a significant increase in \gls{dO} and a significant decrease in \gls{dD} using a \textit{t}-test ($\alpha=\num{0.05}$; Section~\ref{meth:anal:recon}).}} 

\rl{The colors in the image represent the activation amplitude, based on \gls{dEth}, where the amplitude is the difference between stimulus and rest.}
Figure~\ref{fig:funActMap} shows the same relationships in activation amplitude discussed for Figures~\ref{fig:funTraces}\&\ref{fig:exTrace}, with the added caveat that \gls{SD} \gls{phi} displays a larger amplitude than \gls{DS} \gls{I}. Comparing the smallest amplitude to the largest, it is seen that the difference is quite stark with the activation amplitude measured with \gls{DS} \gls{phi} being about three times the one measured with \gls{SD} \gls{I}. Now focusing on the localization of the activation, all data-types found significant activation in the upper central area of the array (with other smaller regions possibly being false positives). \rl{This location approximately corresponds to the primary visual cortex given the array placement in relation to the \gls{Iz}, a cranial landmark of the \textit{occipital} pole, which is the posterior portion of the \textit{occipital lobe}.} Therefore, the upper portion of the array was over the \textit{occipital} lobe, whereas the lower portion was over posterior neck muscles (see Figure~\ref{fig:array}). 
\rl{Finally, the \gls{DS} \gls{phi} map has more white pixels, demonstrating the primary disadvantage of \gls{phi} data, noise, so that more data were eliminated by the methods described in Section~\ref{meth:anal:qual}.} \par

\subsection{\textit{In-Vivo} Systemic Blood Pressure Oscillations}\label{res:osc}
\begin{figure}[bth]
    \begin{center}
        \begin{tabular}{c}
			\includegraphics[width=0.95\linewidth]{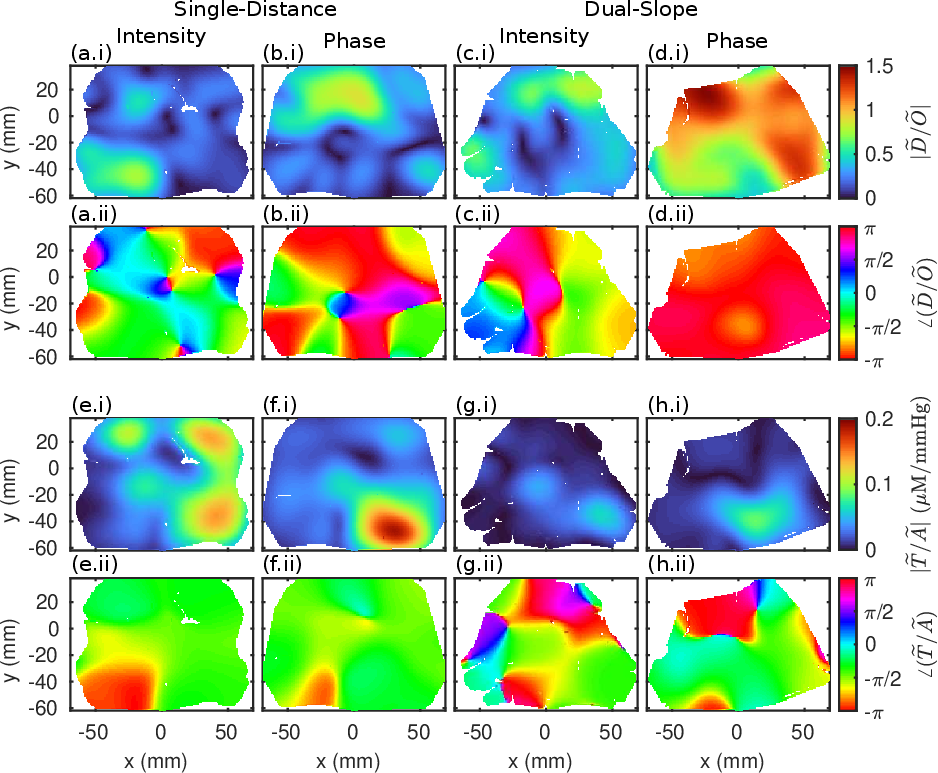}
		\end{tabular}
    \end{center}
    \caption
    {\label{fig:oneFreqMap}
    \Acrfull{DOvec} (a)-(d) and \acrfull{TAvec} (e)-(h) reconstructed maps at \SI{0.1}{\hertz} for \acrfull{SD} \acrfull{I} (a)(e), \acrshort{SD} \acrfull{phi} (b)(f), \acrfull{DS} \acrshort{I}, and \acrshort{DS}\acrshort{phi} (d)(h). White areas indicate a lack of data after data quality restrictions were enforced.}
\end{figure}

\rl{Figure~\ref{fig:oneFreqMap} shows the results from an example of the repeated systemic \gls{ABP} oscillations experiments.}
\q{R1C7b}{\rlA{Methods for image reconstruction using \gls{MPi} to create these images of phasor ratio vectors are described in Sections~\ref{meth:anal:phs}\&\ref{meth:anal:recon}.}} 
The images show either \gls{DOvec} or \gls{TAvec} (Figures~\ref{fig:oneFreqMap}(a)-(d) or Figure~\ref{fig:oneFreqMap}(e)-(h), respectively). Interpretation of these maps requires the simultaneous examination of the amplitude ratio ($|\acrshort{DOvec}|$ or $|\acrshort{TAvec}|$) and the phase difference ($\angle(\acrshort{DOvec})$ or $\angle(\acrshort{TAvec})$) of the phasors (thus the choice of subplot lettering to include (a.i) and (a.ii), for example). This is evident in an image such as Figure~\ref{fig:oneFreqMap}(g) where the upper portion of the image has a $|\acrshort{TAvec}|$ close to zero, making the $\angle(\acrshort{TAvec})$ unreliable and likely dominated by noise. With this guidance for interpretation in-mind, the two different phasor ratio pairs will now be presented in detail.
\q{R2C5b}{\rlB{All results reported here were deemed to represent hemodynamics with significant coherence (Section~\ref{meth:anal:phs}).\cite{blaneyAlgorithmDeterminationThresholds2020}}}
\par

\Gls{DOvec} represents the interplay between \gls{BV} and \gls{BF} oscillations as described by \gls{CHS}\cite{fantiniDynamicModelTissue2014}. 
\rl{When the vector has an angle of $\pi$~\si{rad} and a magnitude of \num{1}, the \gls{Dph} and \gls{Oph} are in-opposition-of-phase and have the same amplitude, thus the measured hemodynamics are dominated by \gls{BF}.}
\rl{On the other-hand, when \gls{DOvec} has an angle of \SI{0}{rad}, the phasors are in-phase, and the measured hemodynamics are dominated by \gls{BV}.}
\gls{DOvec} images are shown in Figure~\ref{fig:oneFreqMap}(a)-(d). All data-types except \gls{SD} \gls{I} found the signature of \gls{BF}-dominated hemodynamic oscillations in the upper portion of the array (corresponding to the \textit{occipital} lobe) but not in the thicker tissue, including skeletal muscle, probed in the lower portion of the array (Figure~\ref{fig:array}). In particular, \gls{DS} \gls{phi} displays \gls{BF}-driven hemodynamics almost everywhere, with a higher amplitude of \gls{DOvec} in the top of the image. Meanwhile, \gls{SD} \gls{phi} and \gls{DS} \gls{I} exhibit \gls{BF}-driven hemodynamics in the upper portion of the image, but very low $|\acrshort{DOvec}|$ values in the lower portion, likely making the $\angle(\acrshort{DOvec})$ in that region unreliable. Finally, the \gls{SD} \gls{I} image is the only one that does not display \gls{BF} domination. This image (Figure~\ref{fig:oneFreqMap}(a)) exhibits a low value of $|\acrshort{DOvec}|$ in the upper and right portions of the imaged area, again indicating unreliability of the $\angle(\acrshort{DOvec})$ image in those regions. However, the lower left portion of the \gls{SD} \gls{I} image does show a larger value of $|\acrshort{DOvec}|$ and corresponding in-phase $\angle(\acrshort{DOvec})$ suggesting a \gls{BV}-driven hemodynamic oscillation or some combination of \gls{BF} and \gls{BV} contributions. In summary, \gls{DS} \gls{phi} measured a \gls{BF} oscillation across the image that is strongest in the upper portion, both \gls{DS} \gls{I} and \gls{SD} \gls{phi} measured a \gls{BF} oscillation in the upper portion of the image, and \gls{SD} \gls{I} measured a \gls{BV} or \gls{BV} mixed with \gls{BF} oscillation in the lower left portion. 
\rl{Note again that the upper portion of the image corresponds to the \textit{occipital} lobe, while the lower portion is likely probing the subject's posterior neck muscles (Figure~\ref{fig:array}).} \par

Now, focus on the \gls{TAvec} images (Figure~\ref{fig:oneFreqMap}). These images relate \gls{BV} oscillations (using the \gls{Tph} as a \gls{BV} surrogate) to the \gls{ABPph}. 
\rl{Any regions of the images with low $|\acrshort{TAvec}|$ are ignored, since they are likely dominated by noise.}
From this, both \gls{DS} \gls{I} and \gls{phi} in the lower right region show similar results with a larger $|\acrshort{TAvec}|$ amplitude  and $\angle(\acrshort{TAvec})$ being mostly in-phase or slightly negative, with \gls{DS} \gls{phi} slightly closer to \SI{0}{rad} than \gls{DS} \gls{I}. This same sort of image is obtained with \gls{SD} \gls{phi}, only with a higher $|\acrshort{TAvec}|$ compared to \gls{DS} \gls{I} and \gls{phi}. Note that since \gls{ABPph} is the same for all images, this means that  \gls{Tph} had a higher amplitude for \gls{SD} \gls{phi} in the lower right than \gls{DS} \gls{I} and \gls{phi}. 
\rl{Moving to \gls{SD} \gls{I}, the observation is made that all regions with a low $|\acrshort{DOvec}|$ show a relatively large $|\acrshort{TAvec}|$.}
\rl{This means that for \gls{SD} \gls{I}, the entire upper portion and lower right portion of the array measured a strong \gls{Oph} amplitude which dwarfed \gls{Dph}, likely due to dominating \gls{BV} oscillations of the arterial vascular compartment.}
Finally, for the \gls{SD} \gls{I} measurement in the lower left portion, there is a low $|\acrshort{TAvec}|$ meaning that the $\angle(\acrshort{TAvec})$ value is likely unreliable. 
\rl{However, these results in combination with the \gls{DOvec} image lead to an additional observation.}
In this portion of the imaged area, \gls{SD} \gls{I} measured \gls{Oph} and \gls{Dph} both with a large amplitude and with an intermediate relative phase \ie{between \SI{0}{rad} and $\pi/2$~\si{rad}}, resulting in a small $|\acrshort{TAvec}|$. This most probably means that \gls{SD} \gls{I} in this portion of the array measured a combination of  \gls{BV} and \gls{BF} contributions to the observed hemodynamics. Finally, it is noted that \gls{SD} \gls{I} appears to be an outlier in that it measured hemodynamics mostly driven by \gls{BV}, whereas \gls{SD} \gls{phi}, \gls{DS} \gls{I}, and \gls{DS} \gls{phi} all found similar results in similar regions and measured hemodynamics more strongly associated with \gls{BF}. \par

\section{Discussion}\label{sec:dis}
\subsection{Dual-Slope Array Distance Design}
\rl{The \gls{DS} array used in this work is extensively described in Reference~\citenum{blaneyDesignSourceDetector2020}, which covers the design methods  to choose \glspl{rho} on the basis of instrumental features and limitations.}
Namely, since the \gls{ISSv2} was considered, the array was designed to include the largest possible $\Delta\acrshort{rho}$ (to improve the signal-to-noise ratio, within the limits imposed by the available dynamic range) and $\acrshort{rho}_{max}$ (to na\"ively increase depth sensitivity, within the limits imposed by the requirement to collect measurable signals). Because of the goal of planar imaging and concerns of dynamic range, the array \glspl{rho} where also homogenized to make all short \glspl{rho} and all long \glspl{rho} the same. From this, the array design featured \glspl{rho} of \SIlist{25;37}{\milli\meter}. Since the work herein shows the first \textit{in-vivo} application of this \gls{DS} array and \gls{DS} \gls{DOI} in general, a more careful examination of how these \glspl{rho} affect the \gls{sen} depth was done via Section~\ref{meth:exp:insil} and presented in Section~\ref{res:insil} and Figures~\ref{fig:violins_meanConst}-\ref{fig:violins_maxConst}. \par

Given the \textit{in-silico} results (Section~\ref{res:insil}), it was found that the chief \gls{rho} parameter affecting \gls{sen} depth was $\bar{\acrshort{rho}}$. 
\rl{This may be somewhat surprising, as one may think that $\acrshort{rho}_{max}$ is the important parameter (this was the na\"ive and incorrect assumption made in Reference~\citenum{blaneyDesignSourceDetector2020}).}
\rl{Additionally, many methods and arrays in \gls{fNIRS} use a very short \gls{rho} \ie{\SIrange{5}{10}{\milli\meter}} for superficial hemodynamics sensing\cite{funaneQuantitativeEvaluationDeep2014,zhangAdaptiveFilteringGlobal2007,gagnonFurtherImprovementReducing2014,saagerMeasurementLayerlikeHemodynamic2008,veesaSignalRegressionFrequencydomain2021} or multiple \glspl{rho} including a short one,\cite{perkinsQuantitativeEvaluationFrequency2021,doulgerakisHighdensityFunctionalDiffuse2019,eggebrechtQuantitativeSpatialComparison2012} which from the perspective of these simulations would hurt \gls{sen} depth since $\bar{\acrshort{rho}}$ would be decreased.}
\rl{However, these methods do not utilize \gls{DS} and instead typically attempt to also measure superficial hemodynamics and then remove them from long \gls{rho} data, which is not the method used here.}
\q{R2C7a}{\rlB{An additional consideration of short or large \gls{rho} measurements,\cite{perkinsQuantitativeEvaluationFrequency2021,doulgerakisHighdensityFunctionalDiffuse2019,eggebrechtQuantitativeSpatialComparison2012} is the difficulty in implementing them. Measurements over a large range of \glspl{rho} where individual sources or detectors are used to collected both short and long \glspl{rho} require large dynamic range or gain switching, which increases instrumental complexity. Furthermore short \gls{rho} measurements are prone to light leakage \ie{detection of light that did not travel through the tissue}. These considerations are not as much of a concern in \gls{DS}, which utilizes two relatively long distances that can be kept within a specified range by proper deign of source-detector arrays.\cite{blaneyDesignSourceDetector2020}}}
\par

\rl{Looking at the \gls{DS} array design (Figure~\ref{fig:array} and Reference~\citenum{blaneyDesignSourceDetector2020}) with these results in mind (Section~\ref{res:insil}), one learns that the depth \gls{sen} of the array may have been improved by lengthening the short \gls{SD} \glspl{rho}.}
\rl{However, reducing $\Delta\acrshort{rho}$ will result in higher \gls{DS} noise\cite{blaneyPhaseDualslopesFrequencydomain2020}.}
\rl{Therefore, this array's \gls{rho} set may actually achieve a good compromise between minimizing noise and maximizing \gls{sen} depth of \gls{DS}, with a $\bar{\acrshort{rho}}$ of approximately \SI{31}{\milli\meter} which is likely the upper end of what is achievable with the \gls{ISSv2} system.} \par

One last result emerging from the \textit{in-silico} simulations is that, on average, \gls{DS} \gls{phi} appears to achieve a better depth sensitivity than \gls{DS} \gls{I} in a two-layer medium, even more so than in the representative homogeneous medium. However, a previously reported two-layer case with a top-layer \gls{mua} of \SI{0.010}{\per\milli\meter} \& \gls{musp} of \SI{1.2}{\per\milli\meter} and bottom-layer \gls{mua} of \SI{0.020}{\per\milli\meter} \& \gls{musp} of \SI{3.0}{\per\milli\meter} reported that \gls{DS} \gls{I} exhibited deeper \gls{sen} than \gls{DS} \gls{phi} for a range of \gls{ztop} \SIrange{4}{10}{\milli\meter}\cite{phamSensitivityFrequencydomainOptical2021a}. 
\rl{That work showed that this was consistent with a small population of \textit{in-vivo} hemodynamics measurements in the human brain.}
From the simulations here, it is seen that this was a special and niche case, as confirmed by examining the \gls{DS} \gls{phi} minus \gls{DS} \gls{I} violin plots (red) in Figure~\ref{fig:violins_meanConst}(b)\&\ref{fig:violins_maxConst}(b), which only show a small tail below zero (below zero meaning \gls{DS} \gls{I} has deeper \gls{sen} than \gls{DS} \gls{phi}). \par

\rl{To find cases that have a similar result to Reference~\citenum{phamSensitivityFrequencydomainOptical2021a} one may look at the negative tail of the difference between \gls{DS} \gls{phi} and \gls{DS} \gls{I} for the \gls{rho} case of \SIlist{25;35}{\milli\meter} in Figure~\ref{fig:violins_maxConst}(b).}
This tail contains \num{1113} of the total \num{16807} combinations of optical properties and layer-thicknesses including the specific case from Reference~\citenum{phamSensitivityFrequencydomainOptical2021a}. 
\rl{Furthermore, all of these cases are ones in which the top-layer \gls{musp} is greater than the bottom-layer \gls{musp}, although these \num{1113} cases are not all of the instances where this scattering relationship is present \ie{a total of \num{7203} cases}.}
For \gls{mua} there is no consistent relationship between the \num{1113} cases and the \glspl{mua} can be found to take the full range of values. \Gls{ztop} however does only take a certain range of values such that only cases where it is \SI{\leq8.3}{\milli\meter} are present in the \num{1113} (for all the varied values of \gls{ztop} the minimum was \SI{5}{\milli\meter} and the next value above \SI{8.3}{\milli\meter} was \SI{10}{\milli\meter}). 
\rl{Therefore, here an extension of the conclusion from Reference~\citenum{phamSensitivityFrequencydomainOptical2021a} is drawn, being that in niche cases \gls{DS} \gls{I} may have deeper \gls{sen} than \gls{DS} \gls{phi} when \gls{musp} is greater in the top-layer for relatively thin \gls{ztop} with little influence from \gls{mua}.} \par

\q{R2C2c}{\rlB{To further investigate specific combinations of two-layer optical properties, an investigation was also done to find cases which recover the true bottom-layer \gls{dmua}. For this investigation, only the \glspl{rho} of \SIlist{25;35}{\milli\meter} were considered. Qualitatively it can seen that most cases which recover the bottom-layer \gls{dmua} belong to \gls{DS} \gls{phi} (Figures~\ref{fig:violins_meanConst}\&\ref{fig:violins_maxConst}). Quantitatively, the number of cases out of \num{16807} can be counted which recover a \gls{dmua} within $\pm\SI{10}{\percent}$ of the true bottom-layer \gls{dmua}. This was $\num{1862}/\num{16807}$ for \gls{DS} \gls{phi}, $\num{364}/\num{16807}$ for the long \gls{SD} \gls{phi}, $\num{29}/\num{16807}$ for the short \gls{SD} \gls{phi}, and $\num{9}/\num{16807}$ for \gls{DS} \gls{I}. Neither long nor short \gls{SD} \gls{I} ever recover values within $\pm\SI{10}{\percent}$ of the true bottom-layer, and furthermore if the threshold was set at $\pm\SI{5}{\percent}$ no \gls{DS} \gls{I} cases would meet the requirement either. To examine which combinations of optical properties result in preferential recovery of the bottom-layer, the \num{1862} cases in which \gls{DS} \gls{phi} reconstructed a value within $\pm\SI{10}{\percent}$ of the true bottom-layer \gls{dmua} are considered next. None of these cases are ones in which the top-layer \gls{musp} is greater than the bottom-layer \gls{musp} \emph{and} the top-layer \gls{mua} is less than the bottom-layer \gls{mua}. However, cases exist in every other scenario. Considering \gls{ztop}, it is noticed that the largest thickness, which meets this requirement, increases as the bottom-layer \gls{mua} is less than the top-layer \gls{mua}. Therefore, it is concluded that the best case scenario is when the top-layer \gls{musp} is less than the bottom-layer \gls{musp} \emph{and} the top-layer \gls{mua} is greater than the bottom-layer \gls{mua}, with the opposite being the worst case.}}\par

Finally, it is noted that due to the negative \gls{sen} regions of \gls{DS} and \gls{phi} data and opposite changes simulated in the top- and bottom-layers, these simulations may artificially give a perceived advantage to \gls{DS} and \gls{phi}. This would occur by a superficial negative \gls{sen} of a negative \gls{dmua} resulting in a positive recovered \gls{dmua} that would be identified as coming from the bottom layer. This effect can been seen in Figure~\ref{fig:violins_meanConst}(a)\&\ref{fig:violins_maxConst}(a) where \gls{DS} \gls{phi} can recover a \gls{dmua} greater than \SI{0.001}{\per\milli\meter}. To confirm that the results and conclusions in this work were not biased by this effect, additional \textit{in-silico} simulations were done with \gls{dmua} only in the top- or only in the bottom-layer. These additional simulations (not presented here for conciseness) confirmed the results and conclusions discussed in this work and show that this possible bias did not affect the conclusions. \par

\subsection{Functional Activation Amplitude Measured with Different Data-Types}
The visual stimulation protocol based on a reversing checkerboard has been extensively studied and utilized in the field of \gls{fNIRS}\cite{uludagCytochromecoxidaseRedoxChanges2004, plichtaEventrelatedFunctionalNearinfrared2006, eggebrechtQuantitativeSpatialComparison2012, bejmInfluenceContrastreversingFrequency2019, reMonitoringHaemodynamicResponse2021}. 
\rl{To name a few, the protocol has been used to study \gls{CCO} redox changes during brain activation\cite{uludagCytochromecoxidaseRedoxChanges2004}, assess the reproducibility of \gls{fNIRS}\cite{plichtaEventrelatedFunctionalNearinfrared2006}, co-register high-density \gls{DOI} and \gls{fMRI}\cite{eggebrechtQuantitativeSpatialComparison2012}, study the effect of checkerboard reversing frequency\cite{bejmInfluenceContrastreversingFrequency2019} (with \SI{8}{\hertz} resulting in the largest amplitude of activation, thus justifying the choice made in this work), and monitor glaucoma patients\cite{reMonitoringHaemodynamicResponse2021}.}
Therefore, in this work neither the protocol nor the measured hemodynamic response is the focus. Instead the visual stimulation protocol has been used as a standard protocol to test and demonstrate the \gls{DS} \gls{DOI} array\cite{blaneyDesignSourceDetector2020}, and compare the various data-types measurable with \gls{FD} \gls{SD} and \gls{DS}. \par

\q{R2C7b}{\rlB{Though it is not the goal of this work, a comparison of these single-subject results can be made against previously published \gls{fNIRS} results which did in-fact intend to study the cerebral hemodynamics and activation location. First, considering the spatial pattern of activation, previous work tends to present bi-lateral regions of activation.\cite{plichtaEventrelatedFunctionalNearinfrared2006,bejmInfluenceContrastreversingFrequency2019} Compared to Figure~\ref{fig:funActMap} it is noticed that only \gls{SD} \gls{phi}, and to some extent \gls{SD} \gls{I}, data shows this pattern, and off-center at that. The off-center nature of the image would imply array mis-alignment \ie{the \gls{Iz} not being the in the center if the array as expected}, which is a possibility. The lack of bi-lateral regions in the \gls{DS} images may be due to a lack of resolution or a loss of data in one region due to noise, this is particularly possible for \gls{DS} \gls{phi}. However, it should be noted that previous results are really only comparable to \gls{SD}, and \gls{SD} \gls{I} in particular, since \gls{DS} is expected to give different \ie{hopefully more brain specific} measurement results. Second, a comparison can be made against previous hemodynamic time traces.\cite{eggebrechtQuantitativeSpatialComparison2012,bejmInfluenceContrastreversingFrequency2019,reMonitoringHaemodynamicResponse2021} Figures~\ref{fig:funTraces}\&\ref{fig:exTrace} show various activation \gls{dO} and \gls{dD} amplitudes for different data-types. Again, previous work should only be compared to \gls{SD} \gls{I} which showed a maximal \gls{dO} amplitude of about \SI{0.4}{\micro\Molar} and a \gls{dD} amplitude of about \SI{-0.2}{\micro\Molar} in Figure~\ref{fig:exTrace}. This is almost exactly the same as the amplitudes reported in Reference~\citenum{eggebrechtQuantitativeSpatialComparison2012}, about half the amplitude reported in the largest response of Reference~\citenum{bejmInfluenceContrastreversingFrequency2019}, and almost the same as the amplitude reported in Reference~\citenum{reMonitoringHaemodynamicResponse2021}. However, these comparisons are tenuous since different methods were used in these works. For example, Reference~\citenum{eggebrechtQuantitativeSpatialComparison2012} reports the traces of a reconstructed voxel based on a \gls{MRI} informed optical property reconstruction prior and  Reference~\citenum{reMonitoringHaemodynamicResponse2021} utilized \gls{TD} and a multi-layer reconstruction. Therefore, despite the general agreement of the amplitudes, a quantitative comparison is not truly possible. Regardless, the goal of this work does not hinge on the results exactly matching previously published works. This is mainly due to two factors. First, that only one subject is presented here since this is a technology development work rather than a study of cerebral dynamics. Second, the array used in this work is rather different than other arrays given that it is designed for \gls{DS}. Even the \gls{SD} results are likely not directly comparable to previous work since the arrangement is rather sparse and contains a small range of distances compared to high density \gls{DOI}.\cite{eggebrechtMappingDistributedBrain2014}}}\par

Comparing the data-types, the higher amplitude response was measured by \gls{DS} \gls{phi}, followed by \gls{SD} \gls{phi}, then \gls{DS} \gls{I}, then \gls{SD} \gls{I} (Figures~\ref{fig:funTraces},\ref{fig:exTrace},\&\ref{fig:funActMap}). This is consistent with the expected sensitivity depth relationships seen in Figures~\ref{fig:violins_meanConst}-\ref{fig:violins_maxConst} and presented in References~\citenum{sassaroliDualslopeMethodEnhanced2019, blaneyPhaseDualslopesFrequencydomain2020, fantiniTransformationalChangeField2019}. One striking result along these lines is that a given pair of optodes may measure no activation response for one data-type but a significant response for another. This was seen in Figure~\ref{fig:exTrace} where a short \gls{SD} \gls{I} trace displayed no activation, but \gls{SD} \gls{phi} from the same pair of optodes did measure a response. 
\rl{Therefore, if one utilizes \gls{FD} \gls{NIRS} the \gls{phi} data should not be ignored since deeper cerebral hemodynamics may be missed by only using \gls{I} data.}
Furthermore, the results also show a further improvement achieved by \gls{DS}, with \gls{DS} \gls{phi} finding the highest amplitude. This, combined with \gls{DS} being insensitive to instrumental drifts and a variety of artifacts\cite{blaneyFunctionalBrainMapping2022,hueberNewOpticalProbe1999}, leads to the recommendation that \gls{DS} is explored more in \gls{fNIRS}. This may be possible with existing data-sets since valid \gls{DS} sets exist in many optical arrays used for \gls{fNIRS}, and these \gls{DS} sets can be found using the methods described in Reference~\citenum{blaneyDesignSourceDetector2020}. \par

\subsection{Spatial Mapping of Hemodynamic Oscillations}
Summarizing the observations made about Figure~\ref{fig:oneFreqMap} in Section~\ref{res:osc}, \gls{SD} \gls{phi}, \gls{DS} \gls{I}, and \gls{DS} \gls{phi} measured similar hemodynamic phasor relationships in similar spatial regions while \gls{SD} \gls{I} retrieved different dynamics. 
\rl{Briefly, all data-types except \gls{SD} \gls{I} found \gls{BF} dominated oscillations in the upper portion of the imaged area, corresponding to the \textit{occipital} lobe (Figures~\ref{fig:array}\&\ref{fig:oneFreqMap}), and found \gls{BV} oscillations in the lower right portion (mostly evident in \gls{SD} \gls{phi} and \gls{DS} \gls{I}), corresponding to neck muscles.}
It is also noted that for \gls{DS} \gls{phi} some mixture of contributions from \gls{BV} and \gls{BF} oscillations may have actually been measured in the lower right portion. Meanwhile, \gls{SD} \gls{I} found \gls{BV} dominated oscillations almost across the entire image except the lower left portion, which measured a combination of \gls{BV} and \gls{BF} oscillations. \par

These results are consistent with previous work using \gls{CHS}, which found that during \gls{ABP} oscillations there is a transition between mainly measuring \gls{BV} oscillations when probing more superficial tissue to measuring more \gls{BF} oscillations when probing deeper tissue, as \gls{BF} changes appear to dominate over \gls{BV} changes in cerebral tissue\cite{khaksariDepthDependenceCoherent2018,blaneyMultidistanceFrequencydomainOptical2019}. It is indeed reasonable to expect that \gls{ABP} oscillations result in stronger \gls{BV} oscillations in scalp tissue than in brain tissue, which is confined within the rigid skull enclosure. This is consistent with previous work considering \gls{CHS} or \gls{DS}\cite{khaksariDepthDependenceCoherent2018,blaneyMultidistanceFrequencydomainOptical2019,blaneyPhaseDualslopesFrequencydomain2020} and may help explain the seemingly unresolved paradoxical results showing cerebral \gls{BV} changes despite the incompressibility and rigidity of the brain fluid dynamic system\cite{kriegerCerebralBloodVolume2012}.
\q{R2C7c}{\rlB{These comparisons to previous works all focus on the phase and amplitude relationships of different data-types, not on the spatial dependence of the hemodynamics shown in Figure~\ref{fig:oneFreqMap}. This is because, to the best of the author's knowledge, spatial mapping of hemodynamic transfer functions measured with \gls{NIRS} and \gls{CHS} has not yet been presented. This makes these results possibly the most novel part of this work. For this reason it is hard to compare the spatial maps presented in Figure~\ref{fig:oneFreqMap} to previously published studies.}}
\par

\rl{One way to examine the spatial mapping results is in terms of the effect of superficial hemodynamics on measurements over the whole array.} 
\rl{Contact pressure between the imaging array and the scalp can affect the superficial hemodynamics\cite{mesquitaInfluenceofprobepressure2013}, and can create inhomogenieties across the array if contact pressure is not consistent throughout the entire array.}
This is particularly a concern when a measurement is dominated by superficial dynamics, since the results may be significantly impacted by the way in which the optical probe is applied on the subject's head. In these studies, no control was put in place to ensure that contact pressure did not suppress superficial hemodynamics, nor was contact pressure measured or controlled. 
\rl{This is one possible explanation for the inhomogeneity in measured amplitude ratio and phase difference for \gls{DOvec} and \gls{TAvec}, especially for \gls{SD} or \gls{I} data-types. Future work will include exploring methods to secure the imaging array with even contact pressure that is monitored and controlled throughout the measurement.} \par

\section{Conclusion}\label{sec:con}
This work sought to investigate and demonstrate \gls{DS} \gls{DOI} via three experiments. The first experiment, which was \textit{in-silico}, examined the \glspl{rho} used in a \gls{DS} set and found that achieving a large mean \gls{rho} is important to maximize depth sensitivity. This first experiment justified the choice of \glspl{rho} used in the \gls{DS} array, which features a mean \gls{rho} of \SI{31}{\milli\meter} for \gls{DS} measurements considering the limitations of the \gls{ISSv2}. Moving from the simulations, the second and third experiments were the first \textit{in-vivo} demonstration of \gls{DS} imaging. The visual stimulation protocol successfully identified the expected activation signal in the primary visual cortex, and the \gls{DS} \gls{phi} data-type recorded the largest amplitude response, indicating its strongest sensitivity to cortical tissue. Then the experiment involving systemic \gls{ABP} oscillations realized spatially resolved measurements of the phase and amplitude relationships between \gls{O} and \gls{D} concentrations, and between \gls{T} concentration and \gls{ABP}. This experiment was also the first work showing spatial mapping of coherent hemodynamic oscillations that are the basis for \gls{CHS}. \rl{In summary, this work reported the first \textit{in-vivo} demonstration of \gls{DS} \gls{DOI}, which aims to apply the intrinsically deeply sensitive \gls{DS} technique to \gls{fNIRS} and \gls{CHS} mapping.} \par

\subsection*{Disclosures}
The authors declare no conflicts of interest.

\subsection*{Acknowledgments}
This research was funded by the \acrfull{NIH} grants R01-NS095334 \& R01-EB029414.

\subsection*{Code, Data, and Materials Availability}
Supporting data and code are avaible at: \href{https://github.com/DOIT-Lab/DOIT-Public/tree/master/DualSlopeImaging}{github.com/DOIT-Lab\\/DOIT-Public/tree/master/DualSlopeImaging} or upon reasonable request.

\glsresetall
\renewcommand{\glossarysection}[2][]{}
\subsection*{Acronyms \& Symbols}
The following acronyms and symbols are used in this manuscript:
\noindent 
\printglossary[type=\acronymtype]
\printglossary[type=symbolslist]

\bibliography{refs}
\bibliographystyle{spiejour}

\subsection*{Biographies} 
\noindent\textbf{Giles~Blaney} is a Postdoctoral Scholar in the \gls{DOIT} lab at Tufts University. He received his \gls{PhD} from Tufts University (Medford, MA USA) in 2022 after working in the same lab with \gls{Prof}~Sergio~Fantini as his advisor. Before that Giles received an undergraduate degree in Mechanical Engineering and Physics from Northeastern University (Boston, MA USA). His current research interests include diffuse optics and its possible applications within and outside of medical imaging.
\vspace{1ex}

\noindent\textbf{Cristianne~Fernandez} is a fourth-year Doctoral Candidate in the \gls{DOIT} lab under the advisement of \gls{Prof}~Sergio~Fantini. Her work mainly focuses on using frequency-domain near-infrared spectroscopy with coherent hemodynamics spectroscopy to measure blood flow oscillations in the brain on healthy human subjects and those in the the neurocritical care unit with a brain injury. Before coming to Tufts University (Medford, MA USA), Cristianne received an undergraduate degree in Biomedical Engineering from Florida International University (Miami, FL USA).
\vspace{1ex}

\noindent\textbf{Angelo~Sassaroli} received a \gls{PhD} in Physics in 2002 from the University of Electro-Communications (Tokyo, Japan). From July~2002 to August~2007, he was Research Associate in the research group of \gls{Prof}~Sergio~Fantini at Tufts University. In September~2007 he was appointed by Tufts University as a Research Assistant \acrlong{Prof}. His field of research is near-infrared spectroscopy and diffuse optical tomography.
\vspace{1ex}

\noindent\textbf{Sergio~Fantini} is \acrlong{Prof} of Biomedical Engineering and Principal Investigator of the \gls{DOIT} at Tufts University. His research activities on the application of diffuse optics to biological tissues resulted in about \num{120} peer-reviewed scientific publications and \num{12} patents. He co-authored with \gls{Prof}~Irving~Bigio (Boston University, Boston, MA USA) a textbook on \say{Quantitative Biomedical Optics} published by Cambridge University Press in 2016. He is a Fellow of the International Society for Optical Engineering (SPIE), Optica, and the American Institute for Medical and Biological Engineering (AIMBE).

\end{spacing}
\end{document}